\documentclass[journal,10pt]{IEEEtran}
\usepackage{cite}
\usepackage{tikz}
\usepackage{amsmath,amsfonts}
\usepackage{array} 
\usepackage{graphicx} 
\usepackage{enumitem} 
\usepackage{enumitem}
\usepackage{tabu,multirow}
\usepackage{algorithmic}
\usepackage{arydshln}  
\usepackage{tabularx}
\usepackage{array}
\usepackage[caption=false,font=normalsize,labelfont=sf,textfont=sf]{subfig}
\usepackage{textcomp}
\usepackage{stfloats}
\usepackage{url}
\usepackage{verbatim}
\usepackage{booktabs}  
\usepackage{array}
\newcolumntype{?}{!{\vrule width 1pt}}
\usepackage{graphicx}
\def\BibTeX{{\rm B\kern-.05em{\sc i\kern-.025em b}\kern-.08em
    T\kern-.1667em\lower.7ex\hbox{E}\kern-.125emX}}
\usepackage{balance}

\begin{document}

\title{ Orthogonal Time Frequency Space for Integrated Sensing and Communication: A Survey}

\author{Eyad~Shtaiwi, Ahmed~Abdelhadi,~Husheng Li, ~Zhu~Han, and H.~Vincent~Poor
\thanks{E. Shtaiwi  is with the Department of Computer and Electrical Engineering and Computer Science, Florida Atlantic University, Boca Raton, FL 33431 USA.}
\thanks{A. Abdelhadi  is with the School of Electrical Engineering \& Computer Science University of North Dakota, Grand Forks, ND 58202, USA.}
\thanks{H. Li is with the School of Aeronautics and Astronautics, and the School
of Electrical and Computer Engineering, Purdue University, USA.}
\thanks{Z. Han is with the Electrical and Computer Engineering Department, University of Houston, Houston, TX 77004, USA, and also with the Department of Computer Science and Engineering, Kyung Hee University, Seoul 446-701, South Korea.}
\thanks{H. Vincent Poor is with the Department of Electrical and Computer Engineering, Princeton University, Princeton, NJ 08544 USA.}
}




\maketitle


\begin{abstract} Sixth-generation (6G) wireless communication systems, as stated in the European 6G flagship project Hexa-X, are anticipated to feature the integration of intelligence, communication, sensing, positioning, and computation. An important aspect of this integration is integrated sensing and communication (ISAC), in which the same waveform is used for both systems both sensing and communication, to address the challenge of spectrum scarcity. Recently, the orthogonal time frequency space (OTFS) waveform has been proposed to address OFDM's limitations due to the high Doppler spread in some future wireless communication systems. In this paper, we review existing OTFS waveforms for ISAC systems and provide some insights into future research. Firstly, we introduce the basic principles and a system model of OTFS and provide a foundational understanding of this innovative technology's core concepts and architecture. Subsequently, we present an overview of OTFS-based ISAC system frameworks. We provide a comprehensive review of recent research developments and the current state of the art in the field of OTFS-assisted ISAC systems to gain a thorough understanding of the current landscape and advancements. Furthermore, we perform a thorough comparison between OTFS-enabled ISAC operations and traditional OFDM, highlighting the distinctive advantages of OTFS, especially in high Doppler spread scenarios. Subsequently, we address the primary challenges facing OTFS-based ISAC systems, identifying potential limitations and drawbacks. Then, finally, we suggest future research directions, aiming to inspire further innovation in the 6G wireless communication landscape.
\end{abstract}

\section{Introduction}
\label{sec:introduction}
Integrated Sensing and Communication (ISAC) is a pivotal technology in the context of the Hexa-X project and the future of wireless networks \cite{hexa1, hexa2, hexa3,6gbeyond, 5gbasic0}. Its paramount importance lies in its capability to address the evolving demands of mobile applications, which include real-time online gaming, vehicle-to-everything (V2X), unmanned aerial vehicle (UAV) communications, and high-speed railway systems \cite{hexa4,hexa5}. These applications necessitate high-speed and ultra-reliable communication, facing challenges in the form of high mobility scenarios and spectrum congestion.

In the forthcoming sixth-generation (6G) wireless communication systems, it is anticipated that ISAC will assume a crucial role in the integration of communication, intelligence, sensing, localization, control, and computation \cite{hexa6,hexa7}. This integration will help ensure the delivery of data at rapid speeds, enhance energy efficiency, and provide accurate localization for both human and Internet of Things (IoT) requirements. By unifying the functions of wireless communication and radar sensing, ISAC optimizes spectrum utilization, mitigates spectrum conflicts, and contributes to reductions in equipment size and energy consumption \cite{drpoor}. Furthermore, ISAC's role in the Hexa-X project underscores its significance as a key technology in the development of the extreme data rates and capacities required for future wireless systems \cite{hexa8}.

The emergence of new applications in the future of wireless communication networks has ushered in a set of fresh challenges that demand immediate attention. While 5G has undoubtedly brought improvements in the quality of service and network access in mobility environments compared to its predecessor, 4G, it is important to address the new requirements of these emerging applications \cite{5g001,5g002,v2x4}. Notably, 5G networks have achieved significant milestones, offering capabilities such as support for high-speed use cases, enhanced energy efficiency, superior spectrum utilization, low latency, high data rates, increased connection density, and enhanced traffic capacity. These achievements have been made possible through the implementation of technologies such as millimeter wave (mmWave) \cite{5gmmwave1, 5gmmwave2, 5gmmwave3,5gmmwave4,5gmmwave5}, ultra-dense networks (UDN) \cite{UDN1, UDN2, UDN3, UDN4, UDN5, UDN6, UDN7}, and massive multiple-input multiple-output (mMIMO) \cite{5gmmimo1,5gmmimo2,5gmmimo3}. \par In mobile wireless environments, the transmitted signal spreads across both time and frequency due to delay spread from multipath and Doppler shift caused by motion \cite{ofdma1,ofdma2}. This results in a doubly dispersive wireless channel, which poses challenges for the conventional orthogonal frequency division multiplexing (OFDM) waveform associated with 5G \cite{ofdma3, ofdma4, ofdma5}. High-mobility scenarios, such as high-speed trains (HSTs) operating at speeds up to 500 km/h \cite{HST1, HST, HST2, HST3, HST4, HST5, HST6, HST7}, vehicle-to-everything (V2X) with terminal speeds reaching 300 km/h \cite{v2x1,v2x2,v2x3}, and unmanned aerial vehicles (UAVs) \cite{uavofdm1, uavofdm2}, introduce rapidly varying channels with high delay and Doppler shifts that exceed the capabilities of traditional OFDM systems.  
\par OFDM relies on dividing the channel bandwidth into narrow-band sub-channels and transmitting data in parallel to achieve high data rates. However, the extreme Doppler shift in high-mobility scenarios leads to significant performance degradation in conventional OFDM systems due to severe inter-carrier interference (ICI) \cite{ofdmici, ICI_in_ISAC2, ICI_in_ISAC, ICI_in_ISAC3} and synchronization challenges \cite{ofdmsynch,ofdmpar}. Additionally, the channel estimation (CE) process \cite{intro_2_channel} and the associated time-frequency adaptation become less effective in these scenarios \cite{otfs_basic0, ofdmici1, ofdmici2}. To address these challenges and improve performance in high-mobility communications, a new 2D modulation scheme called orthogonal time frequency and space (OTFS) has been proposed \cite{otfs_basic1, otfs_intro0, otfs_intro02}. 
OTFS operates in the Doppler-delay (DD) domain, spreading data symbols in this domain using the inverse symplectic finite Fourier transform (ISFFT) and Heisenberg transform at the transmitter, and the SFFT and Wigner transform at the receiver \cite{otfs_intro1, otfs_intro2, otfs_intro3, otfs_intro4, otfs_intro5}. This transformation effectively converts the selective channels into a time-invariant channel in the DD domain, reducing the pilot overhead required for channel estimation \cite{CSI4}. Moreover, the DD domain channel can exhibit sparsity in certain wireless environments, resulting in reduced complexity for channel estimation and data detection \cite{intro_2_channel4, intro_2_channel5, intro_2_channel6, intro_2_channel7}. 
\subsection{Exploring OTFS Modulation for ISAC Systems}
The promising characteristics of OTFS modulation characteristics that make it suitable in high-mobility communications have drawn attention to OTFS as a potential solution for future wireless networks \cite{motivation4}, including the anticipated 6G systems \cite{otfs1111,otfs1112}. OTFS offers advantages such as improved spectral efficiency, latency, robustness in time and frequency, time-frequency localization, and reduced peak-to-average power ratio (PAPR) \cite{10012849,ofdmpar}. Additionally, OTFS enables the integration of radar and communication systems in the same frequency band, addressing the spectrum scarcity challenge. In the context of ISAC systems, OTFS modulation offers a potential solution. OTFS operates in the DD domain, spreading data symbols in this domain instead of the time-frequency domain used by conventional OFDM systems \cite{otfs_basic1, intro_3_spectral2, intro_3_spectral}. This modulation scheme has shown promising characteristics in high-mobility communications, making it suitable for ISAC scenarios \cite{intro_3_spectral3, intro1}. Specifically, OTFS modulation provides several advantages for ISAC systems, as follows:
\begin{enumerate}[
leftmargin=0pt, itemindent=20pt,
labelwidth=15pt, labelsep=5pt, listparindent=0.5cm,
align=left]
    \item \textbf{Robustness in High-Mobility Environments}: OTFS is a modulation and waveform scheme specifically designed to address the challenges posed by highly dynamic and doubly selective channels. In scenarios with high mobility, such as millimeter-wave vehicular ISAC systems \cite{ICI_in_ISAC1}, severe Doppler spreads and ICI can significantly degrade the performance of OFDM in both radar and communications applications \cite{ICI_in_ISAC2, ICI_in_ISAC3, ICI_in_ISAC4}. The lack of guard intervals in the frequency domain allows Doppler shifts caused by moving targets to destroy the orthogonality of OFDM subcarriers at the receiver, resulting in ICI and reduced radar dynamic range \cite{ICI_in_ISAC5, ICI_in_ISAC6}.\par To overcome these challenges, OTFS has been developed as an effective solution. It utilizes joint processing of the time and frequency domains to capture the time-varying characteristics of the channel and mitigate the effects of Doppler spreads and ICI. By doing so, OTFS maintains robust performance even in highly dynamic and doubly selective channels. This makes it particularly suitable for scenarios like ISAC, where severe Doppler spreads and ICI are common. OTFS enables reliable and efficient communication in such environments by effectively handling the challenges associated with highly dynamic channels \cite{ICI_in_ISAC}.
    \item \textbf{Reduced Complexity for Channel Estimation}: In the context of OTFS, the transformation of selective channels into a time-invariant channel in the DD domain plays a crucial role in reducing the complexity of channel estimation \cite{CSI3,intro_2_channel,intro_2_channel5,intro_2_channel6}. This transformation is achieved through the utilization of the  ISFFT and SFFT. By performing this transformation, the time-varying channel characteristics are effectively captured and represented in a more manageable form \cite{intro_2_channel1, intro_2_channel2, CSI_2}. 
 \par This transformation has significant advantages for accurate and efficient channel estimation, which is essential for both sensing and communication tasks within ISAC systems. With the channel transformed into a time-invariant representation \cite{CSI1,CSI6,CSI7}, the complexity of channel estimation is reduced. This reduction in complexity allows for more efficient estimation algorithms to be employed, resulting in improved performance and reduced computational requirements \cite{intro_2_channel2,intro_2_channel6}.

\par Furthermore, the introduction of a low complexity Matching Pursuit (MP) algorithm based on Doppler compensation precoding further enhances the efficiency of the MP algorithm \cite{CSI_1}. This algorithm increases the sparsity of the channel matrix and modifies the stopping criterion, leading to reduced computational complexity while maintaining accurate channel estimation \cite{intro_2_channel4,intro_2_channel7}.

\par Moreover, the use of the Approximate Message Passing (AMP) algorithm for data detection in OTFS introduces another dimension of improvement. An improved version of the AMP algorithm, incorporating covariance pre-processing, has been proposed. This enhanced algorithm further refines the data detection process by leveraging the statistical properties of the channel, ultimately improving the accuracy and reliability of the communication system \cite{intro_2_channel3,intro_2_channel5,intro_2_channel6}.
\par Generally, the combination of the transformation of selective channels in the DD domain, the low complexity MP algorithm, and the improved AMP algorithm with covariance pre-processing contribute to the accurate and efficient channel estimation required in both sensing and communication tasks within ISAC systems. These advancements in channel estimation techniques enable robust and high-performance operation of OTFS in challenging environments \cite{OTFS_dete}.
    \item \textbf{Improved Spectral Efficiency}: OTFS takes advantage of the DD domain to enhance spectral efficiency in ISAC systems. The utilization of the DD domain offers several benefits that contribute to improved spectral efficiency \cite{eFF1,CSI2}. One key advantage of OTFS in the DD domain is the spreading of data symbols \cite{motivation1,motivation2, motivation3}. Instead of using traditional modulation schemes that assign one data symbol to a single sub-carrier, OTFS spreads the data symbols across multiple subcarriers in the DD domain. This spreading effect allows for better utilization of available frequency resources. By distributing the data symbols in the DD domain, OTFS can take advantage of the inherent diversity of the channel and exploit the multipath components more effectively \cite{intro_3_spectral}.

The spreading of data symbols in the DD domain enables higher data rates and increased capacity \cite{10056866}. It allows for the transmission of multiple symbols simultaneously in the frequency domain, resulting in a higher spectral efficiency compared to traditional modulation schemes. With OTFS, more data can be transmitted within the available frequency bandwidth, leading to higher data rates and improved system capacity \cite{intro_3_spectral}.

\par This enhanced spectral efficiency provided by OTFS in the DD domain has significant implications for both sensing and communication applications within ISAC systems \cite{R12}. In sensing applications, where accurate detection and estimation of targets or objects are crucial, the improved spectral efficiency allows for better resolution and discrimination capabilities. It enables the detection of targets with higher accuracy, as well as the identification of multiple targets in a crowded environment \cite{intro_3_spectral2, intro_3_spectral3}.
\par In communication applications, the increased spectral efficiency translates into higher throughput and improved reliability \cite{mMIMO3,R13,mMIMO1,mMIMO2}. The spreading of data symbols in the DD domain mitigates the effects of frequency-selective fading and Doppler shifts, enabling more robust and reliable communication links \cite{9904495}. This is particularly beneficial in ISAC scenarios where severe Doppler spreads and ICI are common challenges \cite{R11}. Therefore, OTFS leverages the DD domain to improve spectral efficiency by spreading data symbols and effectively utilizing available frequency resources. This leads to higher data rates, increased capacity, and improved performance in both sensing and communication applications within ISAC systems \cite{R10}.
    \item \textbf{Compatible with Coexistence Mechanisms}: OTFS modulation offers compatibility with spectrum-sharing mechanisms, making it a valuable tool for the efficient allocation and sharing of frequency resources within ISAC systems \cite{9557830,213131,DSS_OTFS,9903393}. By integrating OTFS with techniques such as time division, frequency division, or code division, the system can dynamically assign different portions of the frequency spectrum to sensing and communication tasks based on their requirements.

In a time division-based approach, OTFS modulation enables the allocation of specific time slots for sensing and communication operations \cite{10152009}. During the sensing time slots, OTFS can be used to transmit signals and capture environmental information, while in the communication time slots, it facilitates data transmission between ISAC devices \cite{10164102,10051983}. This division of time ensures that sensing and communication tasks coexist without significant interference.

Similarly, in a frequency division-based approach, OTFS modulation allows for the allocation of distinct frequency bands for sensing and communication activities \cite{10095505,8647704,9538891}. By utilizing OTFS within the assigned frequency bands, simultaneous sensing and communication operations can take place with minimal interference \cite{9737331, 9737044, 9625452}. This approach optimizes frequency resource usage and enhances overall system efficiency.

Additionally, the integration of OTFS modulation with code division techniques further enhances spectrum sharing within ISAC systems \cite{DSS_OTFS,8647704,9417451}. By employing different spreading codes, OTFS enables efficient separation and discrimination between sensing and communication signals. This capability ensures that the two operations can coexist while minimizing interference and maximizing the utilization of available resources \cite{8892482}.

The compatibility of OTFS modulation with spectrum-sharing mechanisms enables effective and efficient sharing of frequency resources in ISAC systems \cite{9896669,10038836}. It promotes simultaneous sensing and communication operations, ensuring optimal utilization of the spectrum and minimizing interference between tasks. This compatibility significantly enhances the performance and capacity of ISAC systems by enabling seamless coexistence and resource allocation between sensing and communication functions \cite{8108396}.
\end{enumerate}
\par Integrating OTFS modulation into ISAC frameworks requires careful consideration of system design, interference management, and regulatory aspects \cite{work1,work2,work3,work4, work5}. By leveraging the advantages of OTFS in handling high-mobility channels and its compatibility with coexistence mechanisms, ISAC systems can effectively perform both sensing and communication tasks using the same frequency band. Successful implementation of ISAC systems relies on interdisciplinary research, collaboration, and the establishment of regulatory frameworks and standards. By addressing the challenges of interference management, spectrum sharing, system design, and regulatory aspects, we can enable the efficient utilization of spectrum resources and unlock the full potential of integrated sensing and communication applications.
\subsection{Key Insights and Contributions}
\par In this paper, we conduct a comprehensive survey on OTFS-based ISAC systems. We review existing OTFS waveforms for ISAC systems, provide insights into future research directions, and compare the performance of OTFS-enabled ISAC operation with traditional OFDM. We also discuss the main challenges associated with OTFS-based ISAC systems, highlight their drawbacks, and suggest potential future research directions. By exploring the capabilities and limitations of OTFS in ISAC systems, we aim to contribute to the advancement of wireless communication and radar integration in high-mobility scenarios. In this survey paper, we make the following key contributions:
\begin{itemize}[
leftmargin=0pt, itemindent=20pt,
labelwidth=10pt, labelsep=5pt, listparindent=0.5cm,
align=left]
    \item \textbf{Comprehensive Review:} In this comprehensive review, we focus on OTFS-based ISAC systems and provide a detailed analysis of the current state-of-the-art. Our review encompasses existing research works, methodologies, and techniques that are relevant to the integration of radar sensing and communication using the OTFS waveform. This survey is intended to be a resource for both researchers and practitioners who are interested in this field.

The review includes an extensive survey of ISAC system frameworks that make use of the OTFS waveform to achieve simultaneous sensing and communication. We thoroughly analyze and compare various aspects of these systems, such as waveform design, signal processing algorithms, resource allocation strategies, and performance evaluation metrics. By examining these different approaches, we aim to present a comprehensive overview of the field, including the advantages and limitations associated with each approach.

    \item  \textbf{Insights into OTFS Modulation:} We offer insights into the fundamental principles and concepts of OTFS modulation. We explain how OTFS differs from conventional modulation techniques, such as OFDM, and highlight its advantages in high-mobility scenarios with high Doppler spread and time-frequency doubly selective fading channels.
    
    \item  \textbf{Comparison with Conventional OFDM:} We conduct a comparative analysis between OTFS-based ISAC systems and traditional OFDM-based ISAC systems. By highlighting the limitations of OFDM in high-mobility scenarios and the advantages of OTFS in addressing these challenges, we demonstrate the potential performance gains offered by OTFS in joint radar and communication applications. 
    \item   \textbf{Identification of Challenges and Future Research Directions:} We identify and discuss the main challenges and open research directions in OTFS-based ISAC systems. These challenges include synchronization, channel estimation, interference management, coexistence with other wireless systems, and practical deployment considerations. By providing insights into these challenges, we aim to stimulate further research and innovation in this field.
\end{itemize} 
\subsection{Paper Organization}
\begin{figure}[htbp]
    \centering
\includegraphics[width=0.7\columnwidth]{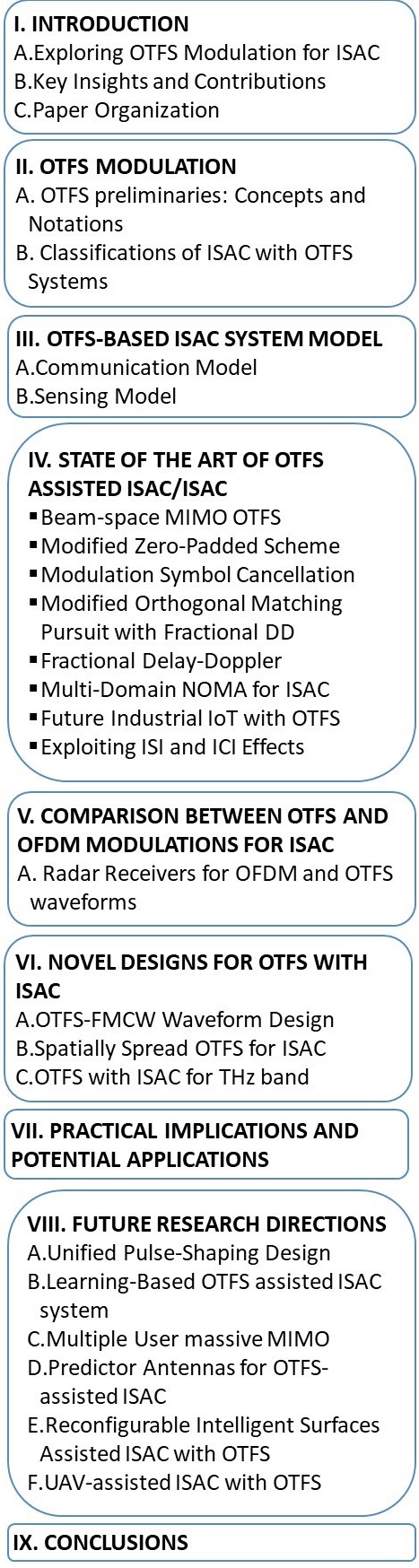}
    \caption{The Structure of the Survey Paper.}
        \label{struct}
\end{figure}

This paper offers a comprehensive analysis of OTFS-based ISAC systems, providing insights into the fundamental principles, a survey of existing frameworks, a comparison with conventional OFDM, an identification of challenges, and an exploration of practical implications and potential applications; see Fig. \ref{struct}. The contributions of this paper aim to facilitate further advances in OTFS-based ISAC systems and promote their adoption in future wireless communication and radar systems.
 \par The rest of the paper is organized as follows. In Section \ref{sec1}, the OTFS modulation principle and ISAC system model are introduced. Section \ref{sec2} describes the OTFS-Based ISAC System Model. Section \ref{sec_state} is a survey on OTFS-based ISAC systems. In Section \ref{otfs_vs_ofdm}, a brief comparison between the OTFS with the conventional OFDM is presented. Section \ref{novel_design} discusses novel designs for OTFS with ISAC systems. Some practical implementation works for ISAC with OTFS are presented in Section \ref{pract}. Section \ref{fut} highlights the challenges and proposes some future research directions. Conclusions are presented in Section \ref{con}. Table \ref{table1} lists the abbreviations used in this paper.

\begin{table}
    \centering
    \caption{List of Abbreviations}
    \begin{tabularx}{\linewidth}{lX}
    \toprule
    \textbf{Abbreviation} & \textbf{Definition} \\
    \midrule
AF & Ambiguity Function \\
AoA & Angle of Arrival \\
AMP & Approximate Message Passing \\
BASE & Base Station \\
CE & Channel Estimation \\
CSI & Channel State Information \\
CP-OTFSM & Cyclic Prefix Orthogonal Time-Frequency-Space Modulation \\
DD & Delay-Doppler \\
DDIPIC & Delay-Doppler Inter-Path Interference Cancellation \\
DoF & Degree of Freedom \\
FFT/IFFT & Fast Fourier Transform/Inverse Fourier Transform \\
HST & High-Speed Trains \\
ICI & Inter-Carrier Interference \\
IoT & Internet of Things \\
ISAC & Integrated Sensing and Communication \\
ISFFT & Inverse Symplectic Finite Fourier Transform \\
LTV & Linear Time-Varying  \\
mMIMO & Massive Multiple-Input Multiple-Output \\
mmWave & Millimeter Wave \\
MP & Matching Pursuit \\
NMSE & Normalized Mean Square Error \\
OFDM & Orthogonal Frequency Division Multiplexing \\
OMPFR & Orthogonal Matching Pursuit with Fractional DD \\
OTFS & Orthogonal Time Frequency and Space \\
OTFSM & Orthogonal Time-Frequency-Space Modulation \\
PAPR & Peak-to-Average Power Ratio \\
PS & Peak-to-Sidelobe Ratio \\
RadTx & Radar Transmitter \\
RF & Radio Frequency \\
RP-OTFSM & Random Padded Orthogonal Time-Frequency-Space Modulation \\
RMSE & Root Mean Square Error \\
RZP-OTFSM & Reduced Zero-Padded OTFSM \\
SDR & Software-Defined Radio \\
SFFT & Symplectic Finite Fourier Transform \\
SNR & Signal-to-Noise Ratio \\
TF & Time-Frequency \\
UAV & Unmanned Aerial Vehicle \\
UE & User Equipment \\
ULA & Uniform Linear Array \\
V2X & Vehicle-to-Everything \\
ZP-OTFSM & Zero-Padded Orthogonal Time-Frequency-Space Modulation \\
ZT & Zak Transform \\
IIoT & Industrial Internet of Things \\
JCAS & Joint Communication and Sensing \\
ML & Maximum Likelihood \\
ISI & Inter-Symbol Interference \\
CP & Cyclic Prefix \\
DFT & Discrete Fourier Transform \\
GLRT & Generalized Likelihood Ratio Test \\
ISLR & Integrated Side Lobe Ratio \\
FIM & Fisher Information Matrix \\
SS-OTFS & Spatially-Spread Orthogonal \\
DFT-s-OTFS & Discrete Fourier Transform and Sparse Orthogonal Time Frequency Space \\
THz & Terahertz \\
Tbps & Terabit-Per-Second \\
RIS & Reconfigurable Intelligent Surface \\
    \bottomrule
    \end{tabularx}
    \label{table1}
\end{table}
\section{OTFS Modulation}
\label{sec1}
As noted above, OTFS modulation is a novel modulation technique that addresses the limitations of traditional OFDM in high-mobility environments \cite{drpoor,ofdm1,10038836}. While both OFDM and OTFS operate in the time and frequency domains, OTFS introduces a new dimension by multiplexing and detecting symbols in the DD domain \cite{10001406, OTFS5, OTFS1}.
In high-mobility scenarios, the received signals experience both frequency and time dispersion due to the multi-path effect and non-negligible Doppler and delay shifts \cite{intro2,otfs1111,intro_2_channel6}. These environments are characterized as doubly-selective channels. Doubly-dispersive channels typically describe outdoor environments where multi-path components are strong, the associated excess delay is large, and both the transmitter and receiver are moving at high velocities \cite{otfs_intro1, otfs_intro02}.

Also as noted above, OTFS has gained attention as a promising waveform for 6G due to its ability to effectively handle high-mobility environments \cite{drpoor}. In particular, by utilizing the DD domain, OTFS can effectively capture and exploit the time-varying characteristics of the channel, making it well-suited for scenarios with strong multi-path components and significant Doppler and delay shifts. The DD domain multiplexing facilitates a thorough characterization of the input-output relationship by exploiting the relatively static channel response, a characteristic not shared by its TF domain counterpart \cite{drpoor}. OTFS has demonstrated good performance in mitigating the challenges posed by doubly selective channels, making it a potential candidate for future communication systems in 6G \cite{otfs_intro0,otfs_basic1}.
\par To delve into the preliminary concepts and notation of OTFS modulation, it is important to understand its distinctive utilization of the DD domain for symbol multiplexing and detection \cite{otfs_basic0}.
\subsection{OTFS Preliminaries: Concepts and Notations}
\begin{figure}
    \centering
    \includegraphics[width=\linewidth]{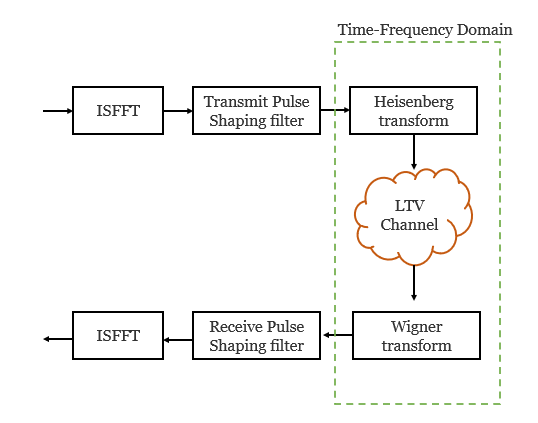}
    \caption{General OTFS Framework.}
    \label{fig:otfs_modulation}
\end{figure}

What differentiates OTFS from the conventional multicarrier waveform is that the former uses the DD domain for multiplexing and detecting the modulation symbols \cite{intro1,otfs_basic1} There are two common implementations of OTFS, namely, the Zak transform (ZT) \cite{drpoor} and the SFFT \cite{otfs_basic1,otfs_intro0,otfs_basic1,otfs_intro02}. For example, the ZT converts the DD symbols into the time domain using two consecutive transformations \cite{intro2,9745430}. Thus, OTFS can be considered as a \textit{pre-} and \textit{post-} processing over a traditional multicarrier system as shown in Fig. \ref{fig:otfs_modulation}.
Let us consider an OTFS system with  $N\times M$ resources in the DD domain. If the time-frequency (TF) plane is sampled at intervals $T$ and $\delta f$, respectively, then the 2D grid is defined as $\Lambda =\lbrace (nT,m\Delta f), n=0,\cdots,N-1, m=0,\cdots, M-1\rbrace$. The $NM$ symbols $x[k, l]$ in the DD domain can be converted into the TF domain as follows:
\begin{equation}X\left[n, m\right]=\frac{1}{\sqrt{MN}}\sum_{k=0}^{N-1}\sum_{l=0}^{M-1}x\left[k, l\right]e^{j2\pi\left(\frac{nk}{N}-\frac{ml}{M}\right)}. \label{dd2tf}\end{equation}
It is worth mentioning that (\ref{dd2tf}) represents the ISFFT \cite{9625231,8686339}. The signals in the time domain are obtained using Heisenberg transform, namely, \begin{equation}x(t)=\sum_{n=0}^{N-1}\sum_{m=0}^{M-1}X[n, m]e^{j2\pi m\triangle f(t-nT)}g_{tx}(t-nT), \end{equation}
where $g_{tx}(t)$ represents a transmitter pulse shaping filter. The time domain received signal after passing the LTV channel is given by \cite{intro3} 
\begin{equation}y(t)=\int_{l/}\int_{\tau}h(\tau, \nu)e^{j2\pi l/(t-\tau)}x(t-\tau)d\tau d\nu+n(t), \end{equation}
where $h(\tau, \nu)$ represents the linear time varying (LTV) channel in DD representation form \cite{8503182,8424569}.
\subsection{Classifications of ISAC with OTFS Systems}
Sensing operations can generally be categorized into two main approaches: active target and passive target \cite{10012421,10164102,9420020}. In passive target sensing frameworks, the sensing target itself is not capable of transmitting and/or receiving sensing signals. However, the sensing target can interact with sensing signals transmitted and/or received by other devices. In this framework, the sensing operation relies on the behavior or characteristics of the sensing targets, such as wireless-based localization. Consequently, the integration of ISAC with OTFS modulation can also be classified into two main types: active target and passive target ISAC with OTFS, based on the role of the sensing target \cite{9747626}.
\subsubsection{Passive Target ISAC with OTFS Framework}
Passive target ISAC with OTFS refers to a  framework in which the the sensing operation does not rely on the active participation of specific sensing targets. In this approach, OTFS modulation is used to sense and communicate with the environment without requiring dedicated devices or cooperation from external entities.

Passive target sensing frameworks typically utilize signals reflected or scattered by objects in the environment. These signals can be used to extract information about the surrounding environment, such as object detection, localization, or tracking. By leveraging the properties of OTFS modulation, passive target ISAC systems can achieve simultaneous sensing and communication capabilities.
The passive target ISAC with OTFS framework has applications in various scenarios. For example, in surveillance systems, OTFS can be used to detect and track moving objects while maintaining communication capabilities. Additionally, in smart environments or IoT applications, passive target ISAC with OTFS enables environment monitoring and data communication without the need for dedicated sensing devices.
\subsubsection{Active Target ISAC with OTFS Framework}
Active target  ISAC with OTFS refers to a framework in which the the sensing operation relies on the active participation of specific sensing targets. These targets are capable of transmitting and/or receiving sensing signals, and OTFS modulation is employed to enable simultaneous sensing and communication.

Active target  ISAC with OTFS systems can be further classified based on different channel topologies and scenarios:

\begin{itemize}
\item {\em Multiple Access Channels with Non-Cooperative Localization}:
In this type of ISAC system, User Terminals (UTs) transmit both sensing and communication information. The Base Station (BS) is responsible for localizing the UTs and decoding their information. OTFS modulation allows for joint processing of the UTs' signals, enabling the extraction of both sensing and communication data. This approach is useful in scenarios such as localization-based applications, where accurate location information of the UTs is required in addition to communication.

\item {\em Broadcast Channel with Non-Cooperative Channel}:

In this scenario, the BS transmits a waveform to the UTs for both sensing and communication purposes. Each UT is capable of eliminating interference, extracting the desired sensing information, and independently communicating its own information. This approach is suitable for scenarios where multiple UTs need to sense the environment and communicate their findings without the need for central coordination. Applications of this framework include distributed sensing and communication systems in dynamic environments.
\end{itemize}

The active target  ISAC with OTFS framework offers flexibility in adapting to various scenarios, enabling joint radar and communication functionalities. It provides the capability to simultaneously sense the environment and communicate, enhancing system efficiency and utilization of resources. The integration of OTFS modulation with active target  ISAC systems opens up opportunities for advanced applications in fields such as autonomous vehicles, smart cities, and intelligent transportation systems.
\section{OTFS-Based ISAC System Model}
\label{sec2}
\begin{figure}[t]
    \centering
    \includegraphics[width= \columnwidth]{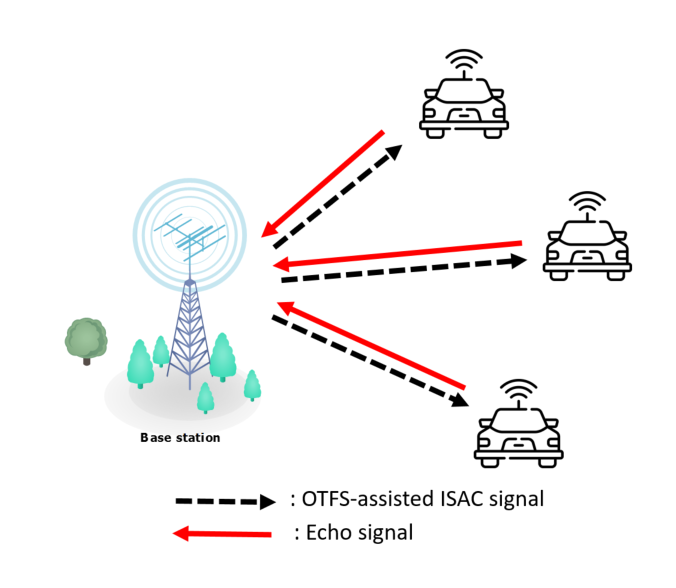}
    \caption{OTFS-assisted ISAC system model}
    \label{fig:systemmodel1}
\end{figure}
Fig. \ref{fig:systemmodel1} depicts a general system model for ISAC assisted with OTFS  which serves an $N_{\rm P}$ UTs \cite{8647394}. The base station (BS) is equipped with a mono-static radar, including a radar transmitter and receiver. The BS is equipped with a uniform linear array (ULA) transmit antenna of $N_{\rm t}$ elements, and a separate ULA receive antenna with $N_{\rm r}$ elements. For ease of implementation, the $N_{\rm P}$ UTs are assumed to be the targets of interest that are particularly applicable to scenarios such as vehicular networks. The transmit and receive arrays are isolated to ensure minimal interference between the sensing echoes and the transmitted signals. Moreover, the isolation guarantees that the sensing echoes received by the received ULA do not disrupt the transmitted signals from the BS.
\par In this system model, the BS is considered a central point for both sensing and communication operations. It utilizes the transmit ULA to send signals and the receive ULA to capture the sensing echoes from the targets. Simultaneously, the targets employ their respective antennas to receive communication signals from the BS. The integration of sensing and communication capabilities in this system model offers several advantages. By leveraging the spatial information captured through the received ULA, the BS can perform accurate sensing and tracking of the targets. This enables various applications, such as target detection, localization, and environment monitoring. Additionally, the targets can effectively receive communication signals from the BS using their multiple antennas, which enhances the system's capacity, coverage, and overall communication performance.
\par Each target of interest, which also serves as a UT, is equipped with $N_{\rm u}$ antennas. These antennas are specifically used for receiving communication information. By employing multiple antennas at the target side, it becomes possible to exploit spatial diversity and achieve better communication performance. The use of multiple antennas can help mitigate fading, combat interference, and improve the overall reliability and quality of the communication links between the targets and the BS.
\par It is worth mentioning that while the system model assumes UTs as targets of interest, the concept can be extended to other scenarios and applications as well. On the other hand, The specific number of antennas at the BS including the transmit and receive ULA, targets can vary depending on the system requirements and design considerations.
\subsection{Communication Model}
In the adopted system model, let us consider a DD grid of $M \times N$ dimension, where $l \in \{0,\ldots, M-1\}
$ and $k \in \{0,\ldots,N-1\}
$ represents the delay and Doppler indices, respectively. The information symbols to the UTs in the DD domain are denoted by $X_{\mathrm{DD}}^{p}\left [{l,k}\right]
$ , where $ p \in \{1,\ldots,N_{\rm P}\}
$. By applying the ISFFT, the DD symbols can be transformed to the TF domain using $X^{p}_{\mathrm{ TF}}\left [{m,n}\right]$. Therefore, the received signal at the $p^{\rm th}$ UT is given by 
\begin{align} \hspace {-2mm}s_{p}(t)=\sum _{n=0}^{N-1}\sum _{m=0}^{M-1}X^{p}_{\mathrm{ TF}}\left [{m,n}\right] g_{\mathrm{ tx}}(t-nT) e^{j2\pi m \Delta f (t-nT)}. {}\end{align}
\par There are two modes of operations at the BS, namely, \textbf{target detection stage} and \textbf{tracking stage}. During the initial target detection stage, the BS detects all relevant targets by generating an omnidirectional or wide-beam signal. However, in the tracking stage, the BS generates beams for both communication and sensing purposes through beamforming.  In other words, the generated multi-beam signals expressed as 
\begin{equation} \tilde { \mathbf {s}}(t) = \mathbf {F} \mathbf {s}(t),\end{equation}
where $\mathbf {F}$ is the beamforming matrix, and $\mathbf {s}(t) = \left [{s_{1}(t),\ldots,s_{P}(t)}\right]^{\mathrm{ T}}
$ is a vector the transmitted signals to all UTs or targets. The columns of the beamforming matrix are used to allocate power and steer the transmitted signal toward the desired directions. To be specific, the $p^{\rm th}$ column of $\mathbf{F}$ represented by  $\mathbf {f}_{p} = \sqrt {\frac {p_{p}}{N_{t}}} \mathbf {a}_{N_{t}}(\tilde {\theta }_{p})
$, where $\mathbf {a}_{N_{t}}(\theta _{p}) = \left [{1,e^{j \pi \sin \theta _{p}},\ldots,e^{j (N_{t}-1)\pi \sin \theta _{p}}}\right]^{\mathrm{ T}}$, steers the transmitted signals to the azimuth direction $\theta _{p}$ and allocates power ${p_{p}}$. 
The communication channel to the $p^{\rm th}$ target can be expressed as 
\begin{equation} \mathbf {H}^{p}_{\mathrm{ DD}}(\tau,\nu) = h_{p} \mathbf {a}_{N_{u}}(\theta _{p}) \mathbf {a}_{N_{t}}^{\mathrm{ H}}(\theta _{p}) \delta (\tau -\tau _{p}) \delta (\nu -\nu _{p}),\end{equation}
where $h_{p}= \sqrt {\frac {c}{4\pi f_{c} d^{2}_{p}}}
$ is the channel gain and $f_c$, $d_p$, $\tau_p$, and $\nu_p$ are the carrier frequency, range, delay and Doppler shift of the $p^th$ UT. The received signal at the $p^{\rm th}$ UT after passing through the multi-carrier communication system is given by \begin{equation}
\begin{aligned}
{y}_{p}(t) = & h_{p} \mathbf{u}^{\mathrm{H}}_{p} \mathbf{a}_{N_{u}}(\theta_{p}) \mathbf{a}_{N_{t}}^{\mathrm{H}}(\theta_{p}) \\
& \times \mathbf{f}_{p} s_{p}(t-\tau_{p}) e^{j 2 \pi \nu_{i} (t-\tau_{p})} + z_{p}(t),
\end{aligned}
\end{equation}

\noindent  and the received OTFS signal is given by \begin{align}
Y^{p}_{\mathrm{DD}}[{l,k}] = & h_{p} \mathbf{u}^{\mathrm{H}}_{p} \mathbf{a}_{N_{u}}(\theta_{p}) \mathbf{a}_{N_{t}}^{\mathrm{H}}(\theta_{p}) \mathbf{f}_{p} \notag \\
& \times X^{p}_{\mathrm{DD}}[{(l-l_{p})_{M},(k-k_{p})_{N}}] + Z^{p}_{\mathrm{DD}}[{l,k}],
\end{align}
where $k_{p} = {\nu _{p}}{N T}
$, $l_{P} = {\tau _{P}}{M \Delta f}$ are integers, and $Z^{p}_{\mathrm{ DD}}\left [{ {l,k} }\right]$ is additive white Gaussian noise.

\subsection{Sensing Model}
Let $\gamma _{p}
$, $\tau_{p}
$, and $\upsilon _{p}$ are the reflected path loss, the delay, and the Doppler shift associated with the $p^{\rm th}$ targets. The sensing channel can be written as 

\begin{equation} \mathbf {H}(t,\tau) = \sum _{p=1}^{P} \gamma _{p} \mathbf {a}_{N_{r}}(\theta _{p}) \mathbf {a}_{N_{t}}^{\mathrm{ H}}(\theta _{p}) \delta (\tau -\tau_{p}) e^{j2\pi \upsilon _{p} t}.\end{equation}
The received echo signals at the BS is given by 
\begin{equation} \mathbf {r}(t) = \sum _{p=1}^{P} \gamma _{p} \mathbf {a}_{N_{r}}(\theta _{p}) \mathbf {a}_{N_{t}}^{\mathrm{ H}}(\theta _{p}) \tilde { \mathbf {s}}(t-\tau_{p})e^{j2\pi \upsilon _{p} t} + \mathbf {w}(t),\end{equation}
where $ \mathbf {w}(t)$ is the noise vector. By neglecting the interference between different targets in the sensing echoes, the BS can effectively distinguish between different targets based on their angles of arrival (AoAs). Consequently, the sensing echo corresponding to the $p^{\rm th}$ target can be extracted from the received signal using a receive beamformer $\mathbf {b}_{p}= \mathbf {a}_{N_{r}}(\tilde {\theta }_{p})
$ which expressed as 
\begin{equation} {r}_{p}(t) = \gamma _{p} \mathbf {b}^{\mathrm{ H}}_{p} \mathbf {a}_{N_{r}}(\theta _{p}) \mathbf {a}_{N_{t}}^{\mathrm{ H}}(\theta _{p}) \mathbf {f}_{p} s_{p}(t-\tau_{p})e^{j2\pi \upsilon _{p} t} + {w}(t).\end{equation}

The Received echo in the DD domain can be written as 
\begin{align} \!\!R^{p}_{\mathrm{ DD}}\left [{ {l,k} }\right]=&G_{a} \sum _{k'=0}^{N-1}\sum _{l'=0}^{M-1} H^{p}_{\mathrm{ DD}}[l',k'] \notag \\&\cdot X^{p}_{\mathrm{ DD}}\left [{ {(l-l')_{M},(k-k')_{N}} }\right] +W^{p}_{\mathrm{ DD}}\left [{ {l,k} }\right],\end{align}
where $G_{a}$ is the composite antenna array gain, $H^{p}_{\mathrm{ DD}}[l',k']
$ is the path loss of the $p^{\rm th}$ target at the DD grid with indices $l'$ and $k'$ corresponding to the delay  $\frac {l'}{M\Delta f}
$ and Doppler $\frac {k'}{N T}
$. The received echos can be expressed in the matrix form by 
\begin{equation} \mathbf {r}^{p}_{\mathrm{ DD}} = G_{a} \mathbf {X}^{p}_{\mathrm{ DD}} \mathbf {h}^{p}_{\mathrm{ DD}} + \mathbf {w}^{p}_{\mathrm{ DD}},\end{equation}


 \section{State of The Art of OTFS assisted ISAC}
 \label{sec_state}
 To gain a comprehensive understanding of the current landscape and advancements in the field of OTFS-assisted ISAC systems, it is important to review the state of the art and recent research developments.
 \subsection{Beam-Space MIMO OTFS} The authors of \cite{work1}  presented a novel approach for MIMO radar in ISAC at mmWave, considering digitally modulated signals. The beam-space MIMO radar design using randomized matrices in the RF domain demonstrates successful target detection and refined parameter estimation. The approach holds potential for applications in various scenarios requiring mmWave ISAC systems. Specifically, the authors investigated the use of ISAC systems operating at mmWave frequency bands. The system consists of one BS equipped with a co-located radar receiver that simultaneously transmits data using an OTFS waveform and performs radar estimation from the backscattered signal. 

 The authors of this paper consider two system function modes: Discovery mode, where a common data stream is broadcast to detect the presence of targets and perform coarse parameter estimation, and Tracking mode, where individual data streams are sent to already acquired users for fine-resolution parameter estimation. The authors proposed the use of multiple randomized reduction matrices to trade off the exploration of the angle of arrival domain and beam pattern directivity, achieving accurate parameter estimation and higher detection probability. They also propose a multi-target generalized likelihood ratio test detector for Discovery mode, while a maximum likelihood-based scheme enables high-resolution estimation in Tracking mode.
A realistic hybrid digital-analog scheme for RF beamforming at mmWave is proposed, considering a reduced number of RF chains compared to the array of antenna elements. The authors of this paper focus on the RF-domain "reduction matrix" of the radar receiver, balancing the exploration capability of the angle domain and the directivity of the beamforming patterns. They propose an efficient maximum likelihood scheme for joint target detection and parameter estimation. Numerical results presented by the authors demonstrate reliable target detection and achievement of the Cramér-Rao lower bound for parameter estimation, confirming the effectiveness of the proposed approach.

This innovative approach capitalizes on multiple randomized reduction matrices to strike an efficient tradeoff between exploration and directivity. By employing a hybrid digital-analog scheme, the system addresses the challenge of having fewer Radio Frequency (RF) chains compared to antenna elements, enabling practical implementation. One of the key strengths of Beamspace MIMO ISAC lies in its successful detection of multiple targets, even when they are separated in at least one dimension. Moreover, the system excels in high-resolution estimation of crucial parameters such as the angle of arrival, delay, and Doppler parameters, which enhances its target detection capabilities.

However, Beamspace MIMO ISAC is optimized for a specific application scenario, primarily tailored for ISAC at mmWave frequencies with a co-located radar receiver.
 This specialization allows for efficient utilization of the technology in the intended context, but it may limit its applicability in other scenarios.
Additionally, there are some inherent complexities associated with the implementation of the Beamspace MIMO ISAC system. The hybrid digital-analog scheme introduces additional intricacies that may impact the overall computational requirements. As with any advanced technology, there is a performance tradeoff to consider. Balancing detection probability and parameter estimation accuracy requires careful consideration to optimize the system's overall performance.
Beamspace MIMO ISAC offers a promising approach to integrated sensing and communication, leveraging randomized reduction matrices and high-resolution parameter estimation to efficiently detect multiple targets in challenging environments. Its practical implementation with a hybrid digital-analog scheme makes it a viable solution, particularly in mmWave frequency scenarios with co-located radar receivers. While the system's capabilities are impressive, careful attention must be given to its inherent complexities and the tradeoffs involved in achieving optimal performance. 

Overall, Beamspace MIMO ISAC represents an exciting advancement in the realm of integrated sensing and communication systems, with the potential to revolutionize various applications in wireless communication, radar, and remote sensing domains. Table \ref{table:work1} presents a summary of the advantages and disadvantages of Beamspace MIMO ISAC with OTFS.
\begin{table}[t]
\centering
\caption{Advantages and Disadvantages of Beamspace MIMO ISAC}
\begin{tabular}{|p{0.4\linewidth}|p{0.4\linewidth}|}
\hline
\multicolumn{2}{|c|}{\textbf{Beamspace MIMO ISAC with OTFS}}\\
\hline
\textbf{Advantages} & \textbf{Disadvantages} \\
\hline
Parameter Estimation: High-resolution estimation of the angle of arrival, delay, and Doppler parameters.  & Specific Application Scenario: Tailored for ISAC at mmWave frequencies with a co-located radar receiver. \\
\hline
Practical Implementation: Hybrid digital-analog scheme addresses the challenge of fewer RF chains compared to antenna elements. & Complexity: Additional complexity in system implementation and potential impact on computational requirements. \\
\hline
Target Detection: Successful detection of multiple targets, even when separated in at least one dimension. & Performance Tradeoff: Balancing detection probability and parameter estimation accuracy may involve tradeoffs. \\
\hline
\end{tabular}
\label{table:work1}
\end{table}

 \subsection{Modified Zero-Padded Scheme}
The work \cite{work2} introduced a new modulation technique called random padded orthogonal time-frequency-space modulation (RP-OTFSM) for ISAC systems. The proposed RP-OTFSM is a modification of the zero-padded scheme (ZP-OTFSM) and offers increased efficiency for OTFSM-based radar systems.

In the paper, it is demonstrated that using a pseudo-random binary sequence with a cyclic prefix as a pilot in the OTFSM transmission system is superior to using an impulse surrounded by zeros, as in the zero-padded OTFSM scheme with an impulsive pilot when the OTFSM signal is intended for sensing purposes as well. The proposed random-padded OTFSM (RP-OTFSM) system was shown to achieve the same or even better bit error rate (BER) compared to the zero-padded OTFSM with an impulsive pilot (ZPP-OTFSM) in fading and high-speed train scenarios/channels. Additionally, the RP-OTFSM system offers the same probability of false alarm detection rate for lower signal-to-noise ratios (SNR) in radar tasks.

However, it is important to note that channel estimation becomes more challenging in the RP-OTFSM method, and the estimated signal obtained from the fast Fourier transform/inverse Fourier transform (FFT/IFFT)-based deconvolution method requires correction. Future work will focus on the practical application of the introduced RP-OTFSM technique using software-defined radio (SDR) in various joint communication and radar/sensing scenarios, particularly considering multi-user and multi-antenna cases.
The paper addresses the challenges posed by the Doppler effect in the traditional multicarrier OFDM systems used in telecommunication services. OFDM-based systems suffer from a loss of sub-carrier orthogonality when transmitters and receivers are in motion, such as in high-speed trains. To overcome this limitation, the authors propose OTFSM as a robust solution for future communication systems.

The OTFSM technique utilizes the Zak transform and offers time-frequency and delay-Doppler interpretations of signal transmission through time-varying communication channels. The OTFSM is an extension of OFDM and introduces extra pre-coding to improve the system's robustness to delay and Doppler effects. The paper explains the basic principles of OTFS modulation and its variants, including reduced zero-padded OTFSM (RZP-OTFSM), zero-padded OTFSM (ZP-OTFSM), reduced cyclic prefix OTFSM (RCP-OTFSM), and cyclic prefix OTFSM (CP-OTFSM).
\begin{table}[t]
\centering
\caption{Advantages and Disadvantages of RP-OTFSM}
\label{table:work2}
\begin{tabular}{|p{0.45\linewidth}|p{0.45\linewidth}|}
\hline
\multicolumn{2}{|c|}{\textbf{Random Padded OTFS (RP-OTFSM) for ISAC systems}} \\
\hline
\textbf{Advantages} & \textbf{Disadvantages} \\
\hline
Improved transmission efficiency & Complexity of implementation \\
\hline
Lower side-lobes in self-ambiguity function (AF) & Channel estimation challenges \\
\hline
Removal of zero fragments & Limited research and development \\
\hline
\end{tabular}
\end{table}

In the context of ISAC systems, the authors propose the ZP-OTFSM with a pilot (ZPP-OTFSM) for channel estimation. The ZPP-OTFSM uses zero padding and includes a pilot signal in the delay-Doppler domain to estimate the channel. However, having zeros in the time-domain waveforms is undesirable for sensing applications. To address this issue, the authors introduce the RP-OTFSM, which replaces zero-padding with random padding. The RP-OTFSM maintains similar bit error rate (BER) characteristics as the ZPP-OTFSM while having lower side lobes in the self-ambiguity function (AF) of the transmitted signal, making it beneficial for sensing tasks.

The  simulation results compared the performance of ZP-OTFSM and RP-OTFSM in terms of BER and AF for a standard fading channel and a high-speed train channel. The RP-OTFSM demonstrates improved efficiency and sensing capabilities compared to the ZP-OTFSM.
One of the primary strengths of RP-OTFSM is its improved transmission efficiency. By utilizing reduced preambles, the system achieves more efficient data transmission, optimizing spectral resources and enhancing overall communication performance.

Furthermore, RP-OTFSM exhibits lower side lobes in its self-ambiguity AF. This characteristic contributes to better interference mitigation and a higher signal-to-noise ratio (SNR), resulting in improved signal detection and reliability. Another advantage of RP-OTFSM lies in the removal of zero fragments, which reduces redundant information and improves spectral efficiency. This removal allows for more efficient use of available bandwidth and enhances the overall system capacity.

However, there are certain challenges associated with RP-OTFSM. The complexity of implementing RP-OTFSM is one such drawback. The reduction of preambles and the optimization of the modulation scheme can introduce additional complexity into the system design and implementation, potentially impacting the practicality and cost of deployment.
Additionally, channel estimation can pose challenges in RP-OTFSM. Accurate channel estimation is crucial for successful communication and can be more challenging with reduced preambles. The system must address these challenges to ensure robust performance in real-world scenarios.
Furthermore, one limitation of RP-OTFSM is the relatively limited research and development compared to more established modulation schemes. As a newer technology, further research and testing are necessary to fully explore its capabilities and address potential limitations.
As summarized in Table \ref{table:work2}, RP-OTFSM offers several advantages, including improved transmission efficiency, lower side-lobes in the self-ambiguity function, and removal of zero fragments for enhanced spectral efficiency. However, the complexity of implementation and challenges related to channel estimation are important considerations. Additionally, as a relatively newer technology, more research and development are required to fully realize the potential of RP-OTFSM and its applicability in various communication scenarios. Overall, RP-OTFSM represents an interesting and promising modulation scheme, and future efforts in advancing its capabilities and addressing its challenges can lead to exciting opportunities for efficient and robust wireless communication systems.
\subsection{Modulation Symbol Cancellation}
An OTFS-based ISAC scheme for high-mobility scenarios like V2X has been proposed in \cite{work3}. The scheme utilizes a mono-static radar to detect targets and simultaneously communicate data to them. At the communication receiver (ComRx) side, the existing OTFS communication system is employed for signal processing. At the radar transmitter (RadTx) side, symbol cancellation is achieved using a high-peak-to-sidelobe-ratio (PSLR) matrix-inversion-based method or an approximate spectrum-division-based method. The proposed scheme allows the estimation of targets’ range and velocity estimates without prior knowledge of targets.  Existing approaches for OTFS-based radar signal processing, such as a matrix-based matched filtering method and an iterative maximum likelihood (ML) method, have their limitations in terms of performance, complexity, and the need for prior knowledge of targets. In this paper, the proposed scheme assumes the use of a mono-static radar for target detection and communication, leveraging the full-guard embedded-pilot OTFS waveform scheme. Channel estimation and symbol detection are performed on the ComRx side, while the RadTx side employs a matrix-inversion-based method or a spectrum-division-based method for symbol cancellation to achieve high-PSLR radar detection. The proposed scheme enables range and velocity estimation of targets without prior knowledge, and the PSLR performance is not determined by the autocorrelation matrix of communication data sequences.
The integrated system discussed in this work combines state sensing and communication functions, leveraging joint radar parameter estimation and OTFS modulation. This integration offers various advantages, including efficient resource sharing, accurate range and velocity estimation, high data rates without compromising radar performance, and robustness to Doppler shifts and sparse channels. However, the system faces challenges due to the complexity of radar parameter estimation algorithms and computation-intensive matrix inversion operations. Spectrum sharing and resource allocation also require careful optimization and soft-output symbol detection for OTFS modulation needs further refinement for optimal performance. Overall, with continued research and development, this integrated approach holds promise for revolutionizing communication and sensing applications, provided that the challenges are effectively addressed to fully unlock its potential.
The integrated system enables seamless blending of state sensing and communication functions, offering benefits such as joint radar parameter estimation, high data rates, and robust OTFS modulation. Nevertheless, challenges concerning complex radar parameter estimation, computational overhead, and optimization of resource sharing and symbol detection must be overcome. As advancements in technology continue, the integrated system's capabilities have the potential to lead to transformative applications in diverse fields, revolutionizing wireless communication, radar systems, and sensing in dynamic environments.
\begin{table}[t]
\centering
\caption{Advantages and Disadvantages of OMPFR}
\label{tab:work5}
\begin{tabular}{|p{0.45\linewidth}|p{0.45\linewidth}|}
\hline
\multicolumn{2}{|c|}{\textbf{Modified Orthogonal Matching Pursuit with Fractional DD}}\\
\hline
\textbf{Advantages} & \textbf{Disadvantages} \\
\hline
Estimates fractional delays and Doppler shifts accurately. & Requires a novel radar channel model. \\
\hline
Lower complexity compared to traditional OMP algorithm for fractional estimation. & Selection of appropriate ISAC system parameters is crucial. \\
\hline
Enables accurate estimation of ranges and velocities. & Limited to OTFS-based ISAC systems. \\
\hline
Improves estimation accuracy with fractional refinement step. & The algorithm assumes targets' delays and Doppler shifts as non-integer multiples of the system resolutions. \\
\hline
\end{tabular}
\end{table}
\subsection{Modified Orthogonal Matching Pursuit with Fractional DD}
The paper presents a novel radar system that combines sensing and communication functionalities using OTFS modulation \cite{work5}. The radar receiver estimates target range and velocity by leveraging transmitted OTFS frames and reflected signals. A new radar channel model is introduced to capture both integer and fractional delays and Doppler shifts caused by targets. The paper proposes a two-step algorithm called modified orthogonal matching pursuit with fractional refinement (OMPFR) to estimate these non-integer delays and Doppler shifts. The algorithm reduces the computational complexity compared to traditional methods by avoiding the need for a large dictionary matrix. Experimental results validate the effectiveness of the OMPFR algorithm in range and velocity estimation. Generally, the paper offers a mono-static OTFS-based radar system for integrated sensing and communication, addressing practical challenges and achieving improved performance.  
OMPFR offers several advantages, including accurate estimation of fractional delays and Doppler shifts, which enhances the system's ability to accurately estimate ranges and velocities. Compared to the traditional OMP algorithm for fractional estimation, OMPFR demonstrates lower complexity, making it a more efficient option for fractional estimation tasks. The algorithm also improves estimation accuracy through its fractional refinement step, further enhancing the overall performance of the system.
However, there are certain drawbacks associated with the OMPFR algorithm. It requires a novel radar channel model, which may add complexity to its implementation and limit its applicability to specific scenarios. The selection of appropriate ISAC system parameters is crucial to achieving optimal performance. OMPFR is also limited to OTFS-based ISAC systems, potentially restricting its use in other modulation schemes. Additionally, the algorithm assumes that targets' delays and Doppler shifts are non-integer multiples of the system resolutions, which might lead to limitations in certain scenarios.
The OMPFR algorithm provides valuable benefits in accurate fractional delay and Doppler shift estimation, facilitating precise range and velocity estimation, Table \ref{tab:work5}. Its lower complexity compared to traditional OMP for fractional estimation makes it an attractive option. However, challenges related to the radar channel model, ISAC parameter selection, and the algorithm's limitation to OTFS-based systems should be considered.
 
\subsection{Fractional Delay-Doppler}
 The authors proposed an algorithm for efficient channel estimation and range/velocity estimation using  OTFS waveform \cite{work6}. The algorithm focuses on canceling inter-path interference (IPI) in the delay-Doppler domain and achieves better performance compared to other channel estimation schemes. It also achieves good root mean square error (RMSE) performance in range and velocity estimation. The algorithm estimates delay and Doppler jointly and includes refinement of estimated channel parameters. The algorithm is applicable in both communication and radar sensing applications. The proposed algorithm, called Delay-Doppler Inter-Path Interference Cancellation (DDIPIC) is proposed which includes Coarse and fine estimation phases utilized to estimate delay and Doppler parameters, and the algorithm runs for a maximum number of iterations. The algorithm's performance in radar parameter estimation, specifically range and velocity estimation, is evaluated and compared to a previous scheme. The proposed algorithm achieves similar performance and is close to the Cramer-Rao lower bound. 
 DDIPIC has several advantages, making it stand out among other channel estimation schemes. It outperforms other methods in channel estimation, achieving accurate and reliable results. The algorithm demonstrates good root mean square error performance in estimating both range and velocity, which are crucial parameters for effective communication and sensing systems. Another advantage of DDIPIC is that it does not assume perfect channel knowledge at the receiver, making it more adaptable to real-world scenarios where channel information may be imperfect or limited.

However, there are certain challenges associated with the DDIPIC algorithm. To effectively use DDIPIC, estimation and cancellation of inter-path interference (IPI) in the Doubly-Dispersive (DD) domain are required. This can add complexity to the implementation and processing of the algorithm. Additionally, DDIPIC assumes perfect knowledge of the number of DD domain paths in the channel, which may not always be the case in practical situations. The complexity of the algorithm increases with the number of paths in the channel, making it more resource-intensive for scenarios with a large number of paths. Furthermore, DDIPIC requires coarse estimation before fine estimation, introducing additional steps and processing overhead in the estimation process.

The DDIPIC algorithm offers advantages such as superior channel estimation performance, accurate range and velocity estimation, and flexibility in handling imperfect channel knowledge. However, the need for IPI estimation and cancellation, perfect knowledge of the number of paths, increased complexity with the number of paths, and the requirement for coarse estimation are important considerations for its practical implementation.

\begin{table}[t]
\centering
\caption{Advantages and Disadvantages of DDIPIC Algorithm}
\label{tab:advantages_disadvantages}
\begin{tabular}{|p{0.45\linewidth}|p{0.45\linewidth}|}
\hline
 \multicolumn{2}{|c|}{\textbf{Delay-Doppler Inter-Path Interference Cancellation
(DDIPIC)}} \\ 
\hline
\textbf{Advantages} & \textbf{Disadvantages} \\
\hline
Outperforms other channel estimation schemes & Requires estimation and cancellation of inter-path interference (IPI) in the DD domain \\
\hline
Achieves good root mean square error performance of range and velocity estimation & Assumes perfect knowledge of the number of DD domain paths in the channel \\
\hline
Does not assume perfect channel knowledge at the receiver & Complexity increases with the number of paths in the channel \\
\hline
Estimates delay and Doppler jointly & Coarse estimation is required before fine estimation \\
\hline
\end{tabular}
\end{table}
\subsection{Multi-Domain NOMA for ISAC}
A proposed scheme called Multi-domain NOMA (MD-NOMA)-based ISAC for B5G and 6G networks is presented in \cite{work7}.  The scheme aims to enable sensing and communication parallelly and non-orthogonally manner in the  TF  and Delay-Doppler DD domains, while also effectively utilizing the additional degree of freedom (DoF) provided by the DD domain. More specifically,  the proposed approach utilizes the TF and DD domains for parallel and non-orthogonal data transfer, while employing an iteration-based receiver to mitigate interference. The scheme demonstrates significant advantages in efficiency and provides a new direction for NOMA and ISAC waveform design. Moreover, The framework presents a new approach to NOMA and ISAC waveform design and holds promise for applications in 6G networks.

The efficiency, achievable rate, and complexity of the proposed MD-NOMA-based ISAC framework are evaluated and discussed. Simulation results demonstrate the effectiveness and significant advantages of the proposed scheme in terms of efficiency. 

The performance of the proposed iteration receiver is evaluated in terms of bit error rate (BER) and sensing performance. The BER curves converge to the ideal curves after seven iterations, indicating impressive performance gains in the initial iterations. The normalized mean square error (NMSE) of delay taps gradually converges, showing the mutual benefit of channel estimation and symbol detection through the iterative process.

Furthermore, the performance of the MD-NOMA-based ISAC scheme is evaluated under different overload factors. The BER performance is studied as the overload rate increases, demonstrating a tradeoff between BER and resource efficiency. The results indicate that a higher overload rate can be tolerated at higher signal-to-noise ratio conditions.  
In terms of efficiency, MD-NOMA-based ISAC offers significant advantages. It demonstrates enhanced resource efficiency, enabling multiple users to access the channel simultaneously, thereby increasing the overall system capacity. However, there may be a possible tradeoff between Bit Error Rate (BER) and resource efficiency, where the system needs to carefully balance these factors to optimize performance.

Regarding achievable rates, MD-NOMA-based ISAC shows better achievable rates compared to baseline systems. This improvement in data transmission rates is a crucial advantage, especially in scenarios with high user demands. However, there is a challenge as the system's performance may degrade with increasing overload rate and signal-to-noise ratio (SNR), which requires careful consideration to ensure reliable communication.

In terms of complexity, MD-NOMA-based ISAC can benefit from complexity reduction using deep learning mechanisms. These mechanisms can optimize system performance and alleviate some of the inherent complexities. However, it's important to note that MD-NOMA inherently has high complexity, which must be managed to ensure feasible implementation.

Regarding multi-user access, there is no specific advantage or disadvantage mentioned in the table, indicating the need for further performance evaluation in multi-user scenarios. The system's capability to support multiple users accessing the channel simultaneously is crucial for its practical applicability.
 \begin{table}[h]
\centering
\caption{Advantages and Disadvantages of MD-NOMA-based ISAC}
\resizebox{\columnwidth}{!}{%
\begin{tabular}{|p{0.4\linewidth}|p{0.4\linewidth}|}
\hline
\multicolumn{2}{|c|}{\textbf{Multi-domain NOMA (MD-NOMA)-based ISAC }}\\
\hline
\textbf{Advantages} & \textbf{Disadvantages} \\
\hline
Significant improvement in efficiency & Possible tradeoff between BER and resource efficiency \\
\hline
Better achievable rate compared to baseline & Performance degradation with increasing overload rate and SNR \\
\hline
Complexity reduction can be done by deep learning mechanisms  & Inherently has high Complexity\\
\hline
- & Performance evaluation for multi-user access needed \\
\hline
\end{tabular}%
}
\end{table}
In general, MD-NOMA-based ISAC offers significant advantages in efficiency and achievable rate, but it requires careful consideration of tradeoffs between BER and resource efficiency. Complexity reduction through deep learning mechanisms can be beneficial, but overall, the system inherently has high complexity that needs to be managed. Additional evaluation in multi-user scenarios and further exploration for combining with traditional PD-NOMA are essential for realizing the full potential of MD-NOMA-based ISAC. With continued research and optimization, this technology holds promise for enabling advanced communication applications in future wireless systems.
\subsection{Future Industrial IoT with OTFS}
The paper \cite{work8} proposed a low-complexity  OTFS sensing method for JCAS in Industrial Internet of Things (IIoT) applications. The method aims to address the challenges associated with effective RF sensing using OTFS waveforms.  
The authors introduced a JCAS as a cost-effective solution that combines communication and sensing functionalities using a single hardware platform and waveform. This approach provides costs, power, and spectrum reductions compared to having separate communication and sensing systems. The authors highlight the suitability of JCAS for IIoT applications, particularly in mission-critical scenarios.
 The authors addressed the challenge of effective RF sensing using OTFS waveforms, which is crucial for JCAS in IIoT applications. In this paper, the authors provided solutions for the existing methods'  limitations in terms of computational efficiency and storage requirements. The authors propose a low-complexity OTFS sensing method that overcomes these limitations and achieves near-maximum likelihood (ML) performance.

The proposed OTFS sensing method includes waveform preprocessing techniques to recover subcarrier orthogonality in the presence of interference and ISI. It also involves the effective removal of communication data symbols in the time-frequency domain without amplifying background noise. An off-grid estimation method is developed for accurately estimating the ranges and velocities of targets.
 The proposed method includes waveform preprocessing techniques to address interference and noise amplification issues, as well as an off-grid estimation method for target parameters.  
 One of the main advantages of this system is its cost-effectiveness. By integrating communication and sensing functionalities, it offers savings in costs, power, and spectrum utilization. Additionally, the system efficiently uses the hardware platform and waveform, maximizing resource utilization.

The low-complexity OTFS sensing method is another advantage, as it achieves near ML performance. The system effectively recovers subcarrier orthogonality, even in the presence of interference and Inter-Symbol Interference (ISI), ensuring accurate estimation of ranges and velocities of targets.

Comprehensive analyses and parameter optimization enhance the system's performance and effectiveness. The robustness of the system across a wide range of velocities is validated through extensive simulations, indicating its potential for practical applications.

On the other hand, the system faces some challenges, such as timing and frequency offset challenges in distributed, noncoherent, asynchronous sensing scenarios. Additionally, there is limited discussion on the networked sensing capabilities of the system, which may be an area for further exploration.

Furthermore, the lack of exploration on integrating sensing results to enhance energy efficiency may be a potential drawback. The proposed system demonstrates several advantages, including cost-effectiveness, efficient resource utilization, low-complexity sensing, and accurate target estimation. It is particularly suitable for mission-critical scenarios in IIoT applications. However, challenges related to timing and frequency offsets, limited discussion on networked sensing, and the need for further exploration of integrating sensing results should be addressed to fully realize the system's potential. With continued research and optimization, this integrated system holds promise for enabling advanced communication and sensing applications, contributing to the development of efficient and reliable wireless systems.
\subsection{Exploiting ISI and ICI Effects}
The authors discussed the use of OTFS as a potential alternative to OFDM in high-mobility communications beyond 5G \cite{R9}. The authors focus on the problem of radar sensing using a joint radar-communications waveform and propose a novel OTFS radar signal model that considers ISI and ICI effects.
The authors demonstrate how ISI and ICI phenomena can be advantageous in surpassing the limitations of existing OFDM and OTFS radar systems in terms of range and velocity detection. They also design a detector/estimator based on a generalized likelihood ratio test (GLRT) that can effectively handle ISI and ICI effects. Simulation results show that embracing ISI/ICI leads to improved detection and estimation performance compared to conventional methods.
The challenges with OFDM at higher operating frequencies are discussed, including PAPR issues, the need for the adaptive cyclic prefix (CP) for combating mobility and fading, and sensitivity to ICI due to Doppler shift in high mobility environments. Existing OTFS approaches for ISAC estimation involve complex operations in the DD domain, requiring iterative interference cancellation processing and suffering from ambiguity in ISI and ICI.
The proposed solutions involve using OTFS waveforms, which have lower PAPR, require less frequent CP adaptation, and can handle larger Doppler shifts. The authors exploit ISI and ICI to address the direction ambiguity in range and velocity estimation, utilizing the time-domain received signal without converting it to the DD domain. They derive a time-domain signal representation for OTFS radar, involving operations such as inverse sparse fast ISFFT, Heisenberg transform (IFFT + pulse shaping), and consideration of CP representation.
The representation of the received backscattered signal in the discrete-time domain involves channel gains, a discrete Fourier transform (DFT) matrix, a frequency-domain steering vector, and a temporal steering vector. The frequency-domain steering vector captures delay-dependent phase rotations resulting in ISI, while the temporal steering vector causes ICI. The authors highlight that ISI can increase the maximum detectable unambiguous delay by a factor of M, and ICI can be exploited to increase the maximum detectable unambiguous Doppler by a factor of $N$.

For the estimation of multiple targets, the authors propose a GLRT detector in the time domain. The detector aims to detect multiple targets and estimate their parameters, including gain, range, and velocity.

Assumptions made in the paper include a CP larger than the round-trip delay, a narrow-band model, and a $K$-tap doubly selective channel.
As future directions, the authors suggest exploring the use of multiple CP in a single OTFS block and evaluating the complexity of the proposed approach compared to DD-domain methods, as the authors claim that their approach has low complexity.
This paper focuses on a joint radar parameter estimation and communication system using OTFS modulation \cite{OTFS_effective}. The scenario is motivated by vehicular applications where a radar-equipped transmitter aims to transmit data to a target receiver while simultaneously estimating parameters such as range and velocity related to the receiver. The receiver is assumed to have been already detected.

The authors explore two perspectives: radar parameter estimation at the transmitter and data detection at the receiver. For radar parameter estimation, they derive an efficient approximated Maximum Likelihood algorithm and Cramér-Rao lower bound for range and velocity estimation. Numerical examples demonstrate that OTFS modulation achieves radar estimation accuracy comparable to state-of-the-art radar waveforms.

In terms of data detection, the authors focus on separate detection and decoding and propose a soft-output detector that leverages channel sparsity in the Doppler-delay domain. The detector's performance is quantified in terms of pragmatic capacity, considering the achievable rate of the channel induced by the signal constellation and the detector's soft output. Simulation results show that the proposed scheme outperforms existing solutions. The paper highlights that an appropriate digitally modulated waveform, such as OTFS, enables efficient joint radar parameter estimation and communication by achieving the full information rate of the modulation and near-optimal radar estimation performance. The authors emphasize that OTFS is well-suited for this purpose.

The following are the underlying assumptions in this paper: a radar-equipped transmitter ready to send data, considers vehicular applications, and employs a low-complexity estimator for continuous values of delay and Doppler. The authors also derive an exact formula for the block-wise input-output relation for OTFS transmission and propose a low-complexity soft-output detector at the receiver. Assumptions include no self-interference, negligible bandwidth expansion due to Doppler, and perfect channel state information at the receiver.
\subsection{Turbo Bi-Static Radar}
The author investigates the concept of OTFS-based joint communication and sensing and discusses the turbo radar sensing algorithm within this framework \cite{turbo}. This approach relies on the intricate dependency of signal propagation and message exchange between the JCS transceiver and communication receiver. The primary objectives include estimating the states of significant reflectors for radar sensing and enabling wireless communication channel estimation.

Specifically, the algorithm begins with statistical motion modeling of reflectors, assuming a known number of significant reflectors and utilizing a linear dynamical system with random perturbations. Both the JCS transceiver and communication receiver remain stationary during this process. The observation mechanism simplifies matters with a linear model for observation and additive noise. Sensing and estimation at both components involve crucial information sharing, such as JCS-target range and perpendicular velocity.

The paper analyzes received signals at both the JCS and communication receivers, detailing the characteristics of these signals for insights into channel modeling in joint communication and sensing scenarios. The incorporation of OTFS signaling enhances the accuracy and realism of the channel representation.

The turbo iteration mechanism, inspired by Turbo decoding in communication systems, is a pivotal aspect. This iterative process involves a dynamic exchange of messages between the JCS transceiver and communication receiver, leveraging historical observations and exploiting temporal correlations. This interaction enhances the system's ability to sense and estimate radar targets, contributing to continual refinement of estimates for increased accuracy in positioning and velocity estimation.

The JCS transceiver and communication receiver use 2-dimensional Fourier transformations to estimate target parameters like position and velocity, with collaborations crucial for accurate estimations. The paper outlines sensing procedures and information fusion at both components.

The turbo sensing and channel estimation process integrate historic prior information using Kalman filtering and turbo iteration, involving the exchange of Gaussian distributions. A lemma for parameter inference under the assumption of a joint Gaussian distribution is introduced, applied to the communication receiver's processing to show a unified prior distribution. numerical simulations results demonstrate the efficacy of the proposed approach, with superior performance in positioning and velocity estimation compared to baseline algorithms.
 \section{Comparison between OTFS and OFDM Modulations for ISAC} 
 \label{otfs_vs_ofdm}
\begin{table}[t]
\caption{Comparison between ISAC-OFDM and ISAC-OTFS}
    \label{tab:isac_comparison}
    \centering
    \begin{tabularx}{\linewidth}{|l|X|X|}
        \hline
        \textbf{Aspect} & \textbf{ISAC-OFDM} & \textbf{ISAC-OTFS} \\
        \hline
        Modulation Type & Time-frequency & Time-frequency delay-Doppler \\
        \hline
        Density & Less & More \\
        \hline
        Doppler Analysis & Difficult & Easier \\
        \hline
        Interference Handling & More (at high Doppler) & Negligible (at high Doppler) \\
        \hline
        Channel Dynamics & Time-invariant & Both time-variant and invariant \\
        \hline
        Pilot Usage & More & Less \\
        \hline
        Location Awareness & Not possible & Possible \\
        \hline
        Sensing Precision & Limited & Improved \\
        \hline
        Range of Values & Moderate & High \\
        \hline
        Fading Resistance & Moderate & High \\
        \hline
        Resource Utilization & Less efficient & More efficient \\
        \hline
        System Complexity & Lower & Higher \\
        \hline
        Adaptability & Limited & Higher \\
        \hline
    \end{tabularx}
\end{table}

 The paper \cite{work9} focused on comparing  OTFS  modulation with OFDM  for ISAC applications. The authors provided the performance analysis of OTFS in terms of parameter estimation and compared it with the widely used OFDM scheme.
the authors provided the Cramér-Rao Lower Bounds (CRLB) for unknown parameter estimation in both SISO and MIMO systems. The CRLB serves as a theoretical benchmark indicating the best possible estimation performance. By analyzing the Fisher Information Matrix (FIM), the authors achieved the CRLB for OTFS and OFDM, considering various system configurations. In \cite{drpoor}, the authors compared the sensing benefits of ISAC in the DD domain to OFDM-based ISAC and argued that DD-ISAC has the potential to outperform OFDM-based ISAC due to its advantage in terms of the ambiguity function. The ambiguity function of DD-ISAC exhibits a robust and 'spike-like' response for the normalized delay and Doppler, indicating superior sensing performance for target detection and parameter estimation using matched filtering.  \par
To provide a comprehensive understanding, the paper presents unified system models for OTFS and OFDM, extending them to MIMO scenarios, Table \ref{tab:isac_comparison} summarizes the distinctions between ISAC-OFDM and ISAC-OTFS frameworks. It discusses the specific differences in the channel matrix expressions between OTFS and OFDM, highlighting the distinctive characteristics of each modulation scheme.
Simulation results are presented to evaluate the fundamental limits of OTFS and OFDM in terms of parameter estimation accuracy. The simulations are conducted in SISO and MIMO setups, considering different key configurations such as  SNR, carrier frequency, and time-frequency resource size. The results demonstrate the advantages of OTFS over OFDM in achieving higher accuracy in parameter estimation for both SISO and MIMO systems.

The paper provided a comprehensive analysis of the performance bounds and potential benefits of OTFS modulation compared to OFDM for integrated sensing and communication applications. It covers theoretical derivations, system modeling, and simulation results, shedding light on the advantages of OTFS in terms of parameter estimation accuracy in various scenarios.
The paper \cite{work9} focused on comparing OTFS modulation with OFDM for ISAC applications. The authors provided the performance analysis of OTFS in terms of parameter estimation and compared it with the widely used OFDM scheme.

The OFDM-based waveform is a good candidate for joint radar and communication systems without affecting the communication system operation. However, ICI will affect the performance of both systems. In other words, if the Doppler shift is above $\frac{\Delta f}{10}$, the orthogonality between subcarriers is not guaranteed. High-mobility environments result in ICI, thus limiting the implementation of OFDM in ISAC. 

Another good candidate for ISAC applications is OTFS, which outperforms the conventional multicarrier system, i.e., OFDM. OTFS uses a shorter CP and is robust against the Doppler shift. 

In this paper, the authors investigated the performance comparison between the OFDM and OTFS receivers under both the TF and DD domain perspectives. The authors introduced three-receiver designs as follows: 
$i)$ Time-domain correlation, which is valid for any multicarrier system but suffers from high computational complexity. 
$ii)$ Symbol-canceling receiver, which divides the TF-domain received signal by the TF-domain communication symbols. 
$iii)$ Delay-Doppler domain receiver. 

The authors compared the receiver designs and the performance of OFDM and OTFS waveforms. In the low-ICI scenarios, where the Doppler shift is less than one-tenth of the frequency spacing, the simulation results showed that different receiver designs peak at the correct range velocity for both OFDM and OTFS. On the other hand, both time-domain and DD-domain receivers reflect the correct range-velocity parameters and are not affected by ICI. 

The results are as follows: 
$i)$ OTFS is \textbf{not} more Doppler tolerant than OFDM under the symbol-canceling receiver design. 
$ii)$ OTFS is more resilient against Doppler shift under the DD domain receiver design at the cost of computational complexities.

The paper provided a comprehensive analysis of the performance bounds and potential benefits of OTFS modulation compared to OFDM for integrated sensing and communication applications. It covered theoretical derivations, system modeling, and simulation results, shedding light on the advantages of OTFS in terms of parameter estimation accuracy in various scenarios. 

Another work proposed a generic description for common multi-carrier radar receivers and compared the sensing performances of two multi-carrier waveforms:  OFDM and  OTFS \cite{work10}. The traditional OFDM is a promising waveform for ISAC systems but suffers from ICI under high Doppler shifts. OTFS modulation, on the other hand, offers improved Doppler tolerance.
\par The authors discussed three different multi-carrier radar receivers available in the literature: time-domain correlation, symbol-canceling receiver, and delay-Doppler domain receiver. Each receiver has its advantages and drawbacks, with varying computational complexity and performance characteristics.

The comparison between the two multi-carrier systems has been conducted for scenarios with low and high ICI. In the low ICI scenario, all receivers perform well, showing peak responses at the correct range and Doppler hypotheses for both OFDM and OTFS waveforms. In the high ICI scenario, the symbol-canceling receiver is unable to resolve the target unambiguously, while the time-domain correlation and delay-Doppler domain receivers provide accurate estimations.

The integrated side lobe ratio (ISLR) is evaluated as a metric to compare the side-lobe characteristics of the receiver-waveform combinations. The results show differences in ISLR between receivers and waveforms, with the symbol-canceling receiver exhibiting increased degradation as Doppler shifts approach the subcarrier separation. The delay-Doppler and correlation receivers demonstrate similar performance, and OFDM shows a lower ISLR in Doppler estimation, while OTFS exhibits a lower ISLR in range estimation with the time-correlation receiver.

 The authors emphasized the radar properties attributed to OTFS and OFDM waveforms are primarily influenced by the associated receivers rather than the waveforms themselves. However, OTFS offers advantages in terms of signal orthogonality for MIMO and opens up possibilities for further research on interference interactions.
\par
\section{Novel designs for OTFS with ISAC}
\label{novel_design}
\subsection{OTFS-FMCW Waveform Design}
 The authors of \cite{work11} proposed a design of a ISAC system using a combination of OTFS and frequency-modulated continuous wave (FMCW) waveforms. ISAC systems aim to map the radio environment while performing communication in the same frequency bands. The proposed design combines the benefits of OTFS for high data rates and FMCW for low-complexity radar reception. By exploiting the simultaneous locality property of FMCW in both time-frequency and delay-Doppler domains, the OTFS and FMCW waveforms are orthogonally superimposed to enable communication and sensing, respectively. The paper provides an analysis of the computational complexity of the proposed design and presents simulation results that demonstrate accurate radar parameter estimation and high data rates.
The paper introduced  a joint OTFS-FMCW waveform design for ISAC systems, aiming to achieve high data rates and low-complexity radar reception. The proposed design combines the benefits of OTFS and FMCW by orthogonally superimposing the waveforms. 
The OTFS-FMCW waveform is a promising design in the field of wireless communication and radar applications, offering several advantages and facing some challenges. On the positive side, it achieves high data rates, making it suitable for applications requiring rapid and efficient data transfer. Additionally, it provides accurate low-complexity radar parameter estimation, making it attractive for radar applications. The waveform's ability to exploit the simultaneous locality property of FMCW in time-frequency and delay-Doppler domains enhances its performance in different scenarios. Furthermore, its capability to perform communication while jointly mapping the radio environment makes it resource-efficient and ideal for high-mobility communication scenarios.

However, the OTFS-FMCW waveform has some drawbacks that need consideration. Its main disadvantage lies in the computational complexity, which could limit its use in IoT devices with restricted power and computational capabilities. Additionally, the waveform may have lower spectral efficiency compared to other waveforms, impacting its performance in crowded frequency bands. To reduce complexity for IoT devices, waveform pre-processing is necessary, potentially adding design and implementation overhead. Moreover, when using multiple chirps for communication, interference issues may arise.

Despite these challenges, the OTFS-FMCW waveform offers significant benefits, such as its low PAPR property and support for long channel coherence time. Furthermore, its compatibility with existing IoT devices makes it an attractive option for upgrades. In conclusion, the adoption of the OTFS-FMCW waveform design depends on striking a balance between its advantages and disadvantages, and careful consideration is needed to determine its suitability for specific applications.
\begin{table}[h]
\caption{Advantages and Disadvantages of the OTFS-FMCW Waveform Design for ISAC.}
\centering
\begin{tabular}{|p{0.45\linewidth}|p{0.45\linewidth}|}
  \hline
  \multicolumn{2}{|c|}{\textbf{OTFS-FMCW Waveform Design for JSAC}}\\
  \hline
  \textbf{Advantages} & \textbf{Disadvantages} \\
  \hline
  Achieves high data rates & Requires computational complexity \\
  \hline
  Provides accurate low-complexity radar parameters estimation & Not suitable for IoT devices with restricted power and computational abilities \\
  \hline
  Exploits simultaneous locality property of FMCW in time-frequency and delay-Doppler domains & May have lower spectral efficiency compared to other waveforms \\
  \hline
  Enables joint mapping of the radio environment while performing communication & Requires waveform pre-processing to reduce complexity for IoT devices \\
  \hline
  Suitable for high mobility communication scenarios & Limited throughput compared to other waveforms \\
  \hline
  Low peak-to average power ratio (PAPR) property & Interference issues when using multiple chirps for communication \\
  \hline
  Supports long channel coherence time & -\\
  \hline
  Compatible with existing IoT devices & -\\
  \hline
\end{tabular}
\label{tab:otfs-fmcw-adv-disadv}
\end{table}

\subsection{Spatially Spread OTFS for ISAC}
The authors proposed the use of spatially-spread orthogonal time-frequency space (SS-OTFS) modulation for ISAC  transmissions \cite{work12}. The paper discussed the advantages of SS-OTFS modulation, which forms beams based on a pre-determined angular grid, in contrast to conventional beamforming that relies on a priori information on angles of departure (AoDs).

The authors derived the input-output relationships for SS-OTFS-enabled ISAC systems in a downlink multi-user MIMO (MU-MIMO) scenario based on delay-Doppler domain channel characteristics. They investigate the angular domain channel features and compare the proposed SS-OTFS scheme with conventional beamforming in terms of signal-to-interference-plus-noise ratio (SINR).  The proposed approach offered robustness to estimation errors on AoDs. The proposed scheme shows improved performance compared to conventional beamforming in terms of SINR for communication and radar. 
SS-OTFS modulation aims to address key challenges in conventional beamforming techniques by forming beams based on a pre-determined angular grid. This unique approach effectively reduces inter-beam interference, enabling efficient spatial focusing of transmitted signals and optimizing the system's overall performance.

One of the notable advantages of SS-OTFS modulation lies in its robustness to estimation errors on angles of departure (AoDs). In real-world scenarios, accurate channel parameter estimation can be challenging due to imperfections and variations, but SS-OTFS demonstrates resilience in such conditions. This robustness ensures reliable communication and sensing, even in complex and dynamic wireless environments.

Moreover, SS-OTFS offers a better Signal-to-Interference-plus-Noise Ratio (SINR) compared to conventional beamforming methods. This translates to improved reception quality and higher interference rejection capabilities, enhancing the reliability and efficiency of the integrated system.

However, like any advanced technology, SS-OTFS modulation also presents certain challenges. Its implementation requires prior knowledge of the delay-Doppler domain channel characteristics, necessitating accurate and up-to-date channel information. This poses a potential limitation in rapidly changing wireless environments where obtaining precise channel data may be difficult.

Additionally, the more complex system design associated with SS-OTFS modulation compared to conventional beamforming demands careful consideration. The added complexity may impact hardware and processing requirements, influencing the feasibility and cost of practical implementation.

Furthermore, the suitability of SS-OTFS modulation may vary based on specific channel conditions and scenarios. Understanding its performance limitations is crucial to effectively harness its benefits and ensure its successful integration into various applications.

Overall, the exploration of SS-OTFS modulation for ISAC transmissions represents an exciting frontier in wireless communication and sensing. The combination of reduced inter-beam interference, robust channel estimation, improved SINR, and efficient spatial focusing holds promise for enhancing the performance and reliability of integrated systems. As researchers and engineers continue to optimize and adapt SS-OTFS to diverse application requirements, this innovative modulation scheme has the potential to revolutionize wireless communication, sensing, and beyond, opening up new opportunities for advanced and efficient technologies in the future.

\begin{table}[bt]
\centering
\caption{Advantages and Disadvantages of SS-OTFS Modulation for ISAC Transmissions}
\begin{tabular}{|p{0.45\linewidth}|p{0.45\linewidth}|}
\hline
\multicolumn{2}{|c|}{\textbf{SS-OTFS Modulation for ISAC Transmissions}}\\
\hline
\textbf{Advantages} & \textbf{Disadvantages} \\
\hline
Forms beams according to a pre-determined angular grid, reducing inter-beam interference. & Requires prior knowledge of the delay-Doppler domain channel characteristics. \\
\hline
Robust to estimation errors on angles of departure (AoDs). & More complex system design compared to conventional beamforming. \\
\hline
Better signal-to-interference-plus-noise ratio (SINR) compared to conventional beamforming. & May have limitations in certain channel conditions or scenarios. \\
\hline
\end{tabular}
\end{table}
\subsection{OTFS with ISAC for THz band}
A new system called DFT-s-OTFS for Terahertz (THz) Integrated Sensing and Communication (ISAC) was introduced in \cite{work13}. The system aims to overcome challenges related to Doppler effects and peak-to-average power ratio (PAPR) in the THz band. It addresses the need for higher spectrum resources in 6G wireless systems and offers ultra-accurate sensing capabilities and Terabit-per-second (Tbps) wireless links.

The proposed DFT-s-OTFS system incorporates a two-stage sensing parameter estimation algorithm. It improves sensing accuracy and velocity estimation by utilizing a coarse estimation method with low computational complexity and a super-resolution estimation method with high accuracy. Additionally, the system enhances power amplifier efficiency by reducing the PAPR compared to existing waveforms like OFDM and DFT-s-OFDM.

The paper highlighted the advantages of THz ISAC, such as its potential applications in vehicle-to-vehicle networks and Terahertz Internet-of-Things (Tera-IoT). It addresses the challenges posed by severe Doppler spread effects in the THz band, particularly in high-mobility scenarios. By equalizing multiple Doppler shifts along each path, DFT-s-OTFS improves link performance in terms of bit error rate (BER) and data rate. Moreover, it addresses the PAPR requirements for THz ISAC systems, ensuring efficient power amplifier operation in the THz band.  The authors demonstrated the effectiveness of the DFT-s-OTFS system in achieving millimeter-level range estimation accuracy and decimeter-level velocity estimation accuracy, outperforming existing waveform technologies in the presence of high-speed targets.
 This innovative approach seeks to address the unique challenges and opportunities presented by THz frequencies, which hold immense promise for high-speed communication and precise sensing applications.

At the core of the DFT-s-OTFS system framework lies a novel technique that utilizes Discrete Fourier Transform and Sparse Orthogonal Time Frequency Space. This combination enables efficient and reliable signal processing, effectively reducing the Peak-to-Average Power Ratio (PAPR) of power amplifiers. This improvement in power amplifier efficiency is critical in energy-conscious THz communication systems, where power consumption is a significant concern.

Moreover, the DFT-s-OTFS system framework leverages a two-stage parameter estimation algorithm, which enhances sensing accuracy and enables decimeter-level velocity estimation. These advancements significantly contribute to more accurate and reliable sensing capabilities, paving the way for sophisticated localization and tracking applications. In fact, the system demonstrates remarkable millimeter-level precision in range estimation, unlocking the potential for a multitude of high-precision applications.

One of the key strengths of the DFT-s-OTFS system lies in its ability to overcome the challenges posed by Doppler effects in high-mobility scenarios. This ensures robust performance in dynamic environments, making it well-suited for scenarios where fast-moving targets or platforms are involved.

However, as with any cutting-edge research, there are certain aspects that need further exploration. While the technical aspects of the DFT-s-OTFS system are compelling, researchers acknowledge the importance of broader discussions to fully understand its potential applications and broader implications. The system's validation through simulations showcases its capabilities, but more extensive experimental validation is warranted to assess its real-world performance and practical implementation.

Despite these challenges, the DFT-s-OTFS system framework represents a promising advancement in THz ISAC. As researchers continue their exploration and optimization, this technology has the potential to revolutionize wireless communication and sensing, unlocking new opportunities for high-frequency applications. By addressing the unique demands of THz frequencies and demonstrating impressive accuracy and efficiency, the DFT-s-OTFS system opens up exciting possibilities for high-speed, high-precision wireless communication and sensing systems in the future. As the journey of research and development continues, the DFT-s-OTFS system framework holds the promise of transforming the way we interact with wireless technologies, ushering in a new era of efficient and reliable THz communication and sensing.

\begin{table}[bt]
\centering
\caption{Advantages and Disadvantages of DFT-s-OTFS system framework}
\begin{tabular}{|p{0.45\linewidth}|p{0.45\linewidth}|}
\hline
\multicolumn{2}{|c|}{\textbf{{OTFS with ISAC for THz band}}}
\\
\hline
\textbf{Advantages} & \textbf{Disadvantages} \\
\hline
Introduces a novel DFT-s-OTFS system framework for THz ISAC. & Focuses primarily on the technical aspects of the proposed system, and may lack broader discussions. \\
\hline
Improves power amplifier efficiency by reducing the PAPR. & Does not include extensive experimental validation, relying mainly on simulations. \\
\hline
Enhances sensing accuracy and velocity estimation with a two-stage parameter estimation algorithm. & Limited discussion on the practical implementation and feasibility of the proposed system. \\
\hline
Demonstrates millimeter-level range estimation and decimeter-level velocity estimation accuracy. & Assumes a THz ISAC context, which may limit applicability to other scenarios. \\
\hline
Addresses challenges related to Doppler effects in high-mobility scenarios. & - \\
\hline
\end{tabular}
\end{table}

 \section{Practical Implications and Potential Applications}
 \label{pract}In this work, the authors demonstrate the successful implementation of the OTFSM waveform in a joint radar-communication system \cite{SDR_OTFS}. The experimental setup uses National Instruments Universal Software Radio Peripheral (USRP) SDR boards for both the transmitter and receiver, allowing them to test the concept in real-world scenarios with static positions. The receiver, equipped with reference and surveillance antennas, serves multiple purposes as a digital data communication decoder and a receiver for sensing purposes.

The study highlights the achievement of two primary objectives: First, after channel estimation and correction, the transmitted data are reliably recovered, indicating the effectiveness of the communication sub-system. Second, the system can detect moving vehicles nearby, showcasing the capability of the radar sub-system.

This research stands as a continuation of the author's previous work, where they simulated OTFSM-based communication and radar-communication systems. The current paper goes beyond simulations and offers concrete experimental evidence to validate the concept.

One significant observation made during the experimental evaluation is the presence of high side-lobes in the OTFS signal's auto-ambiguity function. These side lobes lead to ghost peaks in the calculated range-Doppler maps, which is a challenge that the authors plan to address in future research. They aim to explore signal processing solutions to mitigate this issue and enhance the overall performance of the joint radar-communication system.

Overall, the experimental verification of the proposed OTFSM-based joint radar-communication system shows promising results and opens the door for further advancements in the field. The successful combination of radar and communication functionalities using the OTFSM waveform brings new possibilities for efficient and robust wireless communication systems in complex environments. 
\section{Future Research Directions}
\label{fut}
\subsection{Unified Pulse-Shaping Design}

In the context of Joint Radar-Communication (ISAC) systems integrated with OTFS modulation, future research directions can explore the design and optimization of unified pulse shaping for both the transmitter and the receiver. While existing studies have predominantly focused on using ideal rectangular pulse-shaping filters in their proposed frameworks, there is a need to evaluate the performance of ISAC with OTFS using different pulse shapes.

Currently, there is limited research on assessing the impact of different pulse shapes on the performance of ISAC with OTFS systems. Additionally, the mismatch effect between the pulse shape used at the transmitter and the pulse shape received at the receiver has not been extensively investigated. A comprehensive understanding of these factors and their influence on system performance is crucial for the development of efficient ISAC with OTFS systems.

A future research direction in this area is to explore unified pulse-shaping designs that consider both the transmitter and the receiver. By jointly optimizing the pulse shape at both ends of the communication link, it may be possible to achieve improved system performance and mitigate the effects of pulse shape mismatch. This unified pulse-shaping design can potentially lead to enhanced spectral efficiency, better interference management, and improved overall system performance.

One possible approach for future research is to develop closed-form approximations or analytical models that capture the performance of ISAC with OTFS systems with unified pulse shaping. These models can provide valuable insights into the impact of pulse shaping on various system performance metrics, such as bit error rate, detection probability, or localization accuracy. Such analytical frameworks can guide the design of optimal pulse shapes and facilitate the evaluation of system performance under different scenarios and channel conditions.
In conclusion, the exploration of unified pulse-shaping design for both the transmitter and the receiver is a promising future research direction for ISAC with OTFS systems. By investigating the impact of different pulse shapes and addressing the mismatch effect, researchers can advance the understanding of optimal pulse shaping and develop efficient designs for ISAC with OTFS systems, leading to improved performance and capabilities in future radar-communication integrated systems.
\subsection{Learning-Based OTFS assisted ISAC system}
\par Undoubtedly, the OTFS modulation exhibits remarkable performance when dealing with doubly-disperse channels compared to conventional multi-carrier waveforms. This superiority stems from the distinctive attributes of OTFS channels, particularly their channel sparsity, which effectively reduces computational complexity. Despite these advantages, the existing OTFS works still encounter challenges due to their relatively high computational complexities, making practical implementation in real-world systems difficult.

However, there is a promising solution on the horizon—the integration of a learning-based approach with OTFS in the form of an ISAC with OTFS. This novel direction aims to significantly reduce computational complexity, opening up new possibilities for practical implementations. Surprisingly, despite its potential benefits, no prior research has explored the concept of a learning-based ISAC with OTFS.

Therefore, to harness the full potential of learning-based OTFS-assisted ISAC, considerable research efforts are essential. By investigating and developing this approach further, we can not only pave the way for improved computational efficiency but also enhance the overall performance of OTFS in doubly-disperse channels. Such advancements hold great promise in revolutionizing communication systems and enabling the deployment of efficient, low-complexity OTFS solutions in real-world scenarios. The exploration of learning-based techniques in conjunction with OTFS heralds an exciting future for wireless communication and signal processing research.
\subsection{Multiple User Massive MIMO }
\par The existing OTFS-assisted ISAC considered a simple case scenario in which there is a single communication user with a single antenna and a single target. It is well known that MIMO systems provide an additional spatial DoF. Therefore, MIMO can be exploited to even improve ISAC with OTFS systems in terms of reliability. In addition, multiple communication users scenario needed to be considered in OTFS-assisted ISAC although OTFS assisted MU MIMO ISAC system is a challenging task. Therefore, the OTFS-assisted MU-MIMO ISAC system is another future research direction.
\subsection{Predictor Antennas for OTFS-assisted ISAC}
The Concept of \textit{Predictor Antennas } has been proposed as an energy-efficient and robust downlink data transmission. It is also referred to as a separate receive and training antenna.  The PA is placed on top of a moving terminal, i.e., a vehicle, and aligned in front of one or multiple separate receiving antennas. In general, the terminal is assumed to be moving in  a static EM environment. Due to the movement of the terminals, the receive antenna replaces the PA antenna, and thus, the PA can predict the channel for the receive antenna. The concept can be extended to ISAC systems which allows the UTs to efficiently extract communication information. Future wireless networks are expected to experience a doubly-dispersive nature. Thus, OTFS-empowered ISAC with PA is considered a future research direction. 
\subsection{Reconfigurable Intelligent Surfaces Assisted ISAC with OTFS}
Reconfigurable Intelligent Surfaces (RISs) have emerged as a promising technology for future wireless networks. These man-made surfaces consist of a large number of passive reflecting elements that can manipulate the direction of electromagnetic (EM) waves. Each element of a RIS can induce amplitude and/or phase shifts to the incoming EM signals, allowing for precise control and redirection of the signals in desired directions \cite{RIS1}.

As wireless systems evolve to support high-mobility communications, integrating RISs into these environments presents unique challenges. The interaction between the BS and the RIS, as well as between the RIS and UEs, involves channels that exhibit doubly dispersive characteristics. This doubly dispersive nature arises due to the combination of frequency dispersion caused by mobility and multi-path effects. While existing works \cite{RIS_OTFS1, RIS_OTFS2, RIS_OTFS3, RIS_OTFS4} have proposed RIS-assisted Multiple-Input Multiple-Output (MIMO) systems in high-mobility environments, they have primarily focused on frequency-dispersive channels.

A future research direction is to investigate the integration of RISs in Joint Radar-Communication (ISAC) systems with Orthogonal Time Frequency Space (OTFS) modulation. OTFS is particularly suitable for high-mobility environments due to its ability to effectively handle doubly dispersive channels. By incorporating RISs into the ISAC with OTFS framework, it becomes possible to leverage their reconfigurability to enhance system performance. RIS-assisted ISAC with OTFS can benefit from the capabilities of RISs to steer and shape the transmitted and received signals, optimizing the communication and sensing processes.
The investigation of RIS-assisted ISAC with OTFS systems is expected to address the challenges arising from high-mobility wireless environments. This includes exploring the benefits of RISs in mitigating interference, enhancing signal quality, and improving system capacity. Moreover, it opens up opportunities to optimize the joint design of RIS configuration, OTFS waveform parameters, and ISAC algorithms to achieve enhanced performance and efficiency.
In conclusion, the RISs with OTFS modulation in ISAC systems present an intriguing avenue for future research. By leveraging the reconfigurability of RISs, it becomes possible to explore their potential benefits in high-mobility environments and design innovative solutions for enhanced ISAC with OTFS systems.
\subsection{UAV-assisted ISAC with OTFS}
UAVs have garnered increasing attention due to their flexibility, ease of deployment, high mobility, reliability, and cost-effectiveness. UAV-empowered ISAC have been studied in the literature \cite{UAV1, UAV2}. The UAV's ability to act as both a communication platform and a sensing node offers mutual benefits, enabling communication with multiple ground UTs while simultaneously detecting various ground targets.

However, the existing works on UAV-empowered ISAC have certain limitations. They often assume either time-invariant channels or only consider frequency-selective channels. In the context of future wireless environments characterized by high mobility, these assumptions may not hold, given the doubly-dispersive nature of the channels. As UAVs operate in dynamic scenarios, the proposed works need to be adapted to cope with the challenges posed by time-varying channels with significant delay-Doppler spread.

Therefore, a compelling future research direction is to explore the integration of UAV-empowered ISAC with OTFS. OTFS has demonstrated superior performance in doubly-dispersive channels, making it well-suited for addressing the unique characteristics of high-mobility wireless environments. By incorporating OTFS into UAV-empowered ISAC, we can enhance the system's adaptability, efficiency, and overall performance.

The investigation of UAV-assisted ISAC with OTFS represents a crucial area of research for optimizing wireless communication and sensing in dynamic environments. By overcoming the challenges posed by high-mobility scenarios and leveraging the advantages of OTFS, this integrated approach has the potential to revolutionize UAV applications in communication, sensing, and beyond. Advancements in this field will pave the way for innovative solutions in aerial robotics, wireless networks, and remote sensing domains.
\section{Conclusions}
\label{con}
In this survey paper, we present a comprehensive analysis of OTFS-based  ISAC systems. Our review encompasses existing research works, methodologies, and techniques relevant to the integration of radar sensing and communication using the OTFS waveform. The survey provides insights into the fundamental principles of OTFS modulation, highlighting its advantages in high-mobility scenarios with challenging time-frequency doubly selective fading channels. Additionally, we conduct a comparative analysis between OTFS-based ISAC systems and traditional OFDM-based ISAC systems, demonstrating the potential performance gains of OTFS in joint radar and communication applications. Furthermore, we identify and discuss key challenges and open research directions in OTFS-based ISAC systems, including synchronization, channel estimation, interference management, coexistence with other wireless systems, and practical deployment considerations. Our contributions aim to advance the understanding and development of OTFS-based ISAC systems. The survey serves as a valuable resource for researchers and practitioners interested in this field, providing valuable insights and encouraging further research and innovation. By addressing the challenges and leveraging the advantages of OTFS modulation, we promote the adoption of OTFS-based ISAC systems in future wireless communication and radar applications, ultimately enhancing their performance and applicability in real-world scenarios.

 
\end{document}